\documentclass[conference]{IEEEtran}
\IEEEoverridecommandlockouts
\usepackage{graphicx}
\usepackage[skip=2pt]{caption}
\usepackage{subcaption}
\usepackage{cite}
\usepackage{epsfig}

\usepackage{inconsolata}

\begin{document}

\captionsetup[figure]{font=small, labelfont={bf},labelformat={default},labelsep=period,name={Figure }}
\captionsetup[table]{font=small, labelfont={bf},labelformat={default},labelsep=period,name={Table }}
%\captionsetup[subfigure]{labelfont=bf,textfont=normalfont,singlelinecheck=off,justification=raggedright}
%\captionsetup[lstlisting]{labelfont={bf},labelformat={default},labelsep=period,name={Listing }}

\title{Transaction-level Model Simulator for Communication-Limited Accelerators}

%\title{AccTLMSim: Transaction-Level Modeling Based Accelerator Simulator for Convolutional Neural Networks}

\author{\IEEEauthorblockN{Sunwoo Kim, Jooho Wang, Youngho Seo, Sanghun Lee, Yeji Park, Sungkyung Park\IEEEauthorrefmark{1} and Chester Sungchung Park}
\IEEEauthorblockA{\textit{Department of Electrical Engineering, Konkuk University, \IEEEauthorrefmark{1}Department of Electronics Engineering, Pusan National University}\\
E-mail: \{sunwkim, joohowang, younghoseo, sanghunlee, yejipark, chester\}@konkuk.ac.kr, \IEEEauthorrefmark{1}fspark@pusan.ac.kr}\\\vspace*{-0.8cm}
}

\date{}
\maketitle

\thispagestyle{empty}

\begin{abstract}

Rapid design space exploration in early design stage is critical to algorithm-architecture co-design for accelerators. In this work, a pre-RTL cycle-accurate accelerator simulator based on SystemC transaction-level modeling (TLM), AccTLMSim, is proposed for convolutional neural network (CNN) accelerators. The accelerator simulator keeps track of each bus transaction between accelerator and DRAM, taking into account the communication bandwidth. The simulation results are validated against the implementation results on the Xilinx Zynq. Using the proposed simulator, it is shown that the communication bandwidth is severely affected by DRAM latency and bus protocol overhead. In addition, the loop tiling is optimized to maximize the performance under the constraint of on-chip SRAM size. Furthermore, a new performance estimation model is proposed to speed up the design space exploration. Thanks to the proposed simulator and performance estimation model, it is possible to explore a design space of millions of architectural options within a few tens of minutes.

\end{abstract}

\section{Introduction}

Nowadays convolutional neural networks (CNNs) have emerged as a promising solution in computer vision [1]. The use of custom hardware accelerators helps to improve energy-efficient implementation of CNNs [2]-[10]. In particular, most of the state-of-the-art CNN accelerators focus on reducing the DRAM accesses that account for a main portion of energy consumption. The well-known on-chip buffering, which allows for storing and reusing a subset of data, is an efficient way to reduce DRAM energy consumption. To cope with the on-chip SRAM size limitation, the loop tiling is often incorporated into the on-chip buffering [2]-[4], [11], [12]. The performance of a CNN accelerator is often determined by the communication bandwidth, which tends to be limited by either DRAM latency or bus protocol overhead [3], [11].

Algorithm-architecture co-design, which is essential for effective accelerator design, often necessitates an extensive design space exploration, and there have been many reports of design space exploration for CNN accelerators [2]-[5], [9],[11]-[13]. Recalling that today’s system-on-a-chip (SoC) tends to take thousands of engineer-years to implement, it is of primary importance to make design decisions as early in the design flow as possible. Several pre-RTL accelerator simulators are proposed to make it feasible to explore the design space in the early design step [10], [14]-[16]. NVDLA [10] provides a SystemC-based simulation model for a parameterizable CNN accelerator. gem5-Aladdin [14] supports the automated conversion of a high-level language algorithm description into the accelerator architecture in addition to full-system power-performance simulations. However, they lack the ability to model the accelerator on a low-level, thereby making it hard to evaluate the relevant performance impact of architectural parameters. Although PARADE [15] includes the detailed models of DRAM, the accelerator is connected to the rest of SoC through the custom network-on-chip (NoC) instead of the standardized system bus such as AMBA-compliant crossbar. Synopsys Platform Architect [16] provides a wide variety of SystemC-based low-level simulation models for DRAM subsystems and on-chip buses, but it does not include any low-level simulation model for accelerators.

There have been numerous reports on the performance evaluation of CNN accelerators [2], [3], [8], [9], [11]-[13], [17]. Although some of them consider the performance impact of DRAM latency [3], [17], the communication bandwidth is assumed to be static, for example, independent of DRAM access pattern. Furthermore, the performance impact of bus protocol overhead has never been investigated. This may lead to an incorrect design decision since the communication bandwidth is in practice dynamic due to DRAM latency and bus protocol overhead, as will be shown in this paper.
To the best of our knowledge, this is the first publication that evaluates the impact of DRAM latency and bus protocol overhead on a CNN accelerator. Specifically, the no local reuse (NLR), the well-known dataflow, is assumed to be connected to off-chip DRAM through AXI bus. In order to facilitate the performance evaluation, we propose a pre-RTL cycle-accurate accelerator simulator based on SystemC transaction-level modeling (TLM) [18] and validate it using the Xilinx Zynq. Using the proposed simulator, the accelerator is optimized to maximize the performance for the given on-chip SRAM size. It is shown that both DRAM latency and bus protocol overhead should be taken into consideration when the tile sizes of loop tiling are chosen. In addition, a new performance estimation model, is proposed to speed up the design space exploration. Finally, based on the design space exploration results, a new layer-dependent loop tiling is proposed to provide an additional performance gain.

\section{CNN Accelerator}

This section describes the accelerator dataflow assumed in this work, which is followed by the overall system including a DRAM controller and system buses. In addition to the accelerator dataflow, this section describes the data movement outside the accelerator, i.e., from/to the off-chip DRAM through the system buses.

\setcounter{figure}{0}
\captionsetup[figure]{font=small, labelfont={bf},labelformat={default},labelsep=period,name={Listing }}
\begin{figure}[t] %%% t: top, b: bottom, h: here
\begin{center}
\includegraphics[width=.8\linewidth]{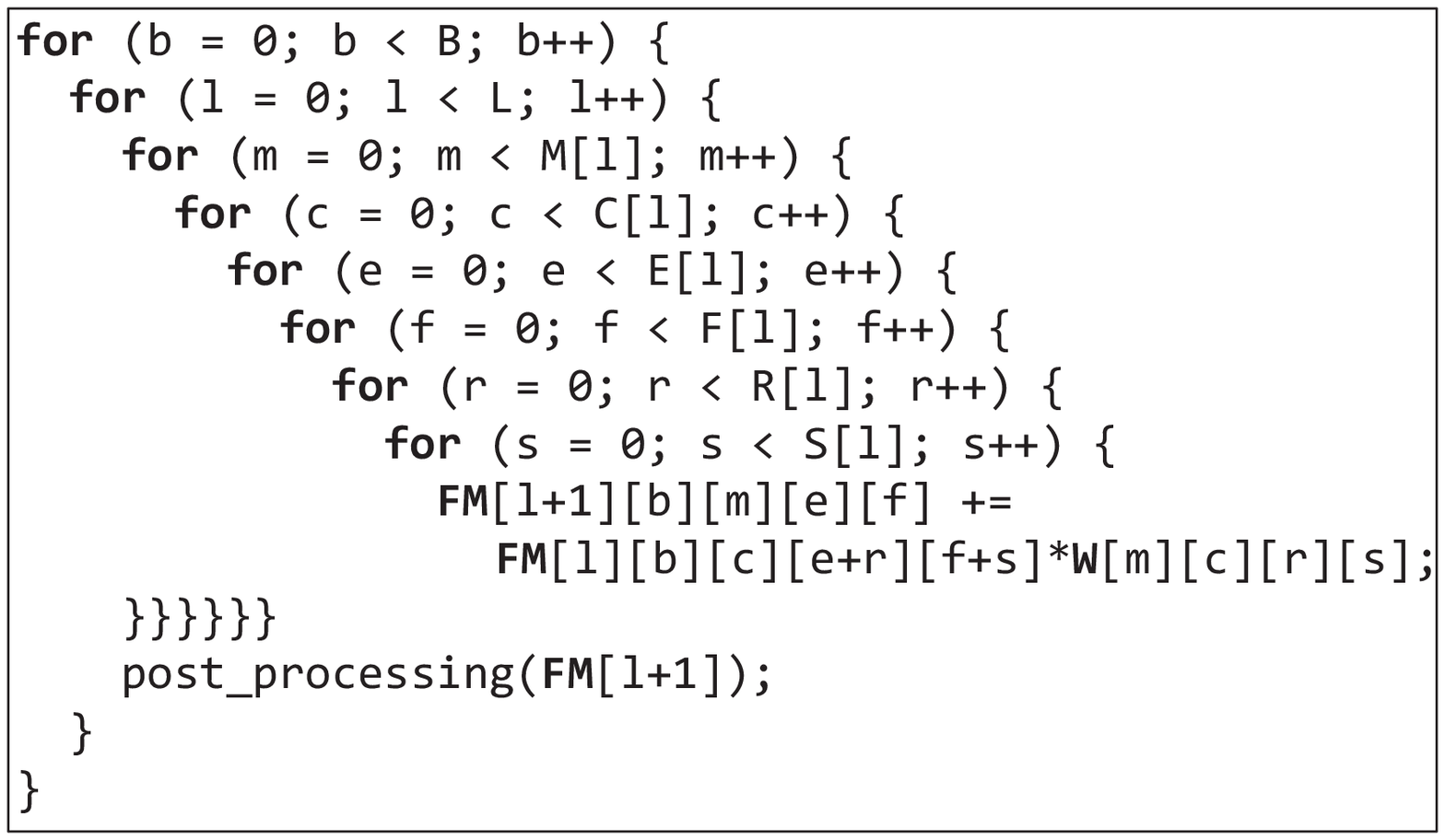}
\end{center}
%\captionof{lstlisting){-}
\caption{Pseudo code of convolutional layers}
\label{lst:1}
\end{figure}

\subsection{Accelerator Dataflow}

Convolution is the main operation in CNN that involves a number of multiply-and-accumulate (MAC) operations. Each convolutional layer can be expressed as eight levels of loops, as shown in Listing 1. For each of B images, C input feature maps of the l-th layer (FM[l]) are convolved with M filters (W), generating M output feature maps of the same layer (i.e., one for each filter), or equivalently,  M input feature maps of the (l+1)-th layer (FM[l+1]). The height and width of an input feature map, a filter and an output feature map are denoted by H x W, R x S and E x F, respectively.

The data movement within an accelerator, which is referred to as dataflow, should be optimized to improve the performance and energy efficiency of CNN accelerators [8].  No local reuse (NLR) is one of the well-known dataflows, which is characterized as a matrix multiplication using a set of reduction trees [2]-[5]. On-chip buffering is a commonly used technique to reduce off-chip DRAM access. In order to fit into on-chip SRAM, it is often combined with the loop tiling that partitions the data into smaller chucks. The detailed operation of NLR is captured in Listing 2 where FM and W denote the feature maps and filters stored in the off-chip DRAM, respectively. In Listing 2, the input feature maps, filters and output feature maps stored in the on-chip buffers are denoted by \texttt{\textbf{ibuf}}, \texttt{\textbf{wbuf}} and \texttt{\textbf{obuf}}, respectively. The loop tiling is parameterized by the tile sizes for batch (TB), output feature map (TE, TF), input channel (TC) and output channel (TC) that are all layer-dependent. Similarly, the input feature maps, filters and output feature maps stored in the local registers are denoted by \texttt{\textbf{ix}}, \texttt{\textbf{wx}} and \texttt{\textbf{ox}}, respectively. Once the on-chip buffers are loaded with a set of input feature maps and filters (along the loops across tiles), the MAC array takes TC input pixels and TC$\times$TM weights and then returns TM partial sums back to the on-chip buffers (along the loops within a tile). Upon completion of the convolution over a tile, the resulting output feature maps are stored into the off-chip DRAM. This procedure is referred to as a processing pass and each convolutional layer is divided into multiple processing passes. The feature maps and filters processed by each processing pass must be stored within the on-chip buffers and thus the on-chip buffer size is determined by the tile sizes. The loops across tiles except the two outer-most loops (i.e., the loops over \texttt{\textbf{e}}, \texttt{\textbf{f}}, \texttt{\textbf{m}} and \texttt{\textbf{c}}) represent inter-pass iterations whereas the loops within a tile except the two inner-most loops (i.e., the loops over \texttt{\textbf{r}}, \texttt{\textbf{s}}, \texttt{\textbf{tb}}, \texttt{\textbf{te}} and \texttt{\textbf{tf}}) represent intra-pass iterations. The two inner-most loops (i.e., the loops over \texttt{\textbf{tm}} and \texttt{\textbf{tc}}) are unrolled to exploit the hardware parallelism. In general, the unroll factors UM and UC are constrained to be equal to the tile sizes TM and TC, respectively [2]. In this case, these tile sizes are no longer layer-dependent since the unroll factors are mostly layer-independent, i.e., uniform across different convolution layers [2], [4]. The same constraint is assumed throughout this paper, unless otherwise mentioned.

\begin{figure}[t] %%% t: top, b: bottom, h: here
\begin{center}
\includegraphics[width=.92\linewidth]{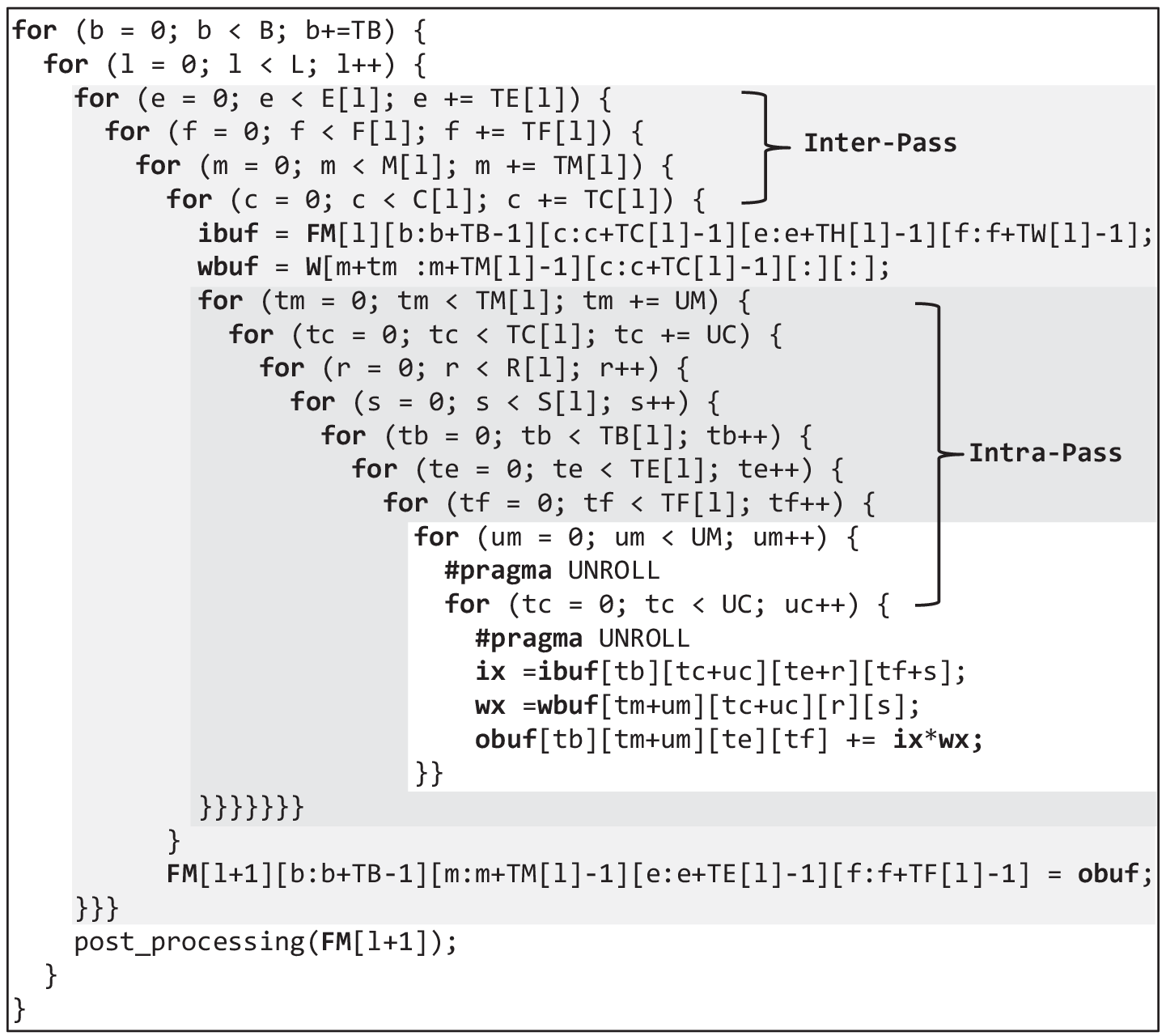}
\end{center}
%\captionof{lstlisting){-}
\caption{Pseudo code for NLR with loop tiling}
\label{lst:1}
\end{figure}

Figure 1 is the block diagram of an accelerator with TC = 3 and TM = 4. The MAC array consists of a total of TC$\times$TM MAC units and it is capable of performing the same number of MAC operations each cycle. We assume a scratchpad-based accelerator that is interfaced to the external system by direct memory access controllers (DMACs) through on-chip buffers implemented by SRAM. Double buffering is adopted to decouple communication from computation by allowing one buffer to be accessed by the DMACs and another by the MAC array. A pair of on-chip buffers together with a DMAC is dedicated to each of the data types. The DMACs are responsible for transferring data from/to the off-chip DRAM through the system bus. In addition, a network-on-chip (NoC) is used to connect the MAC array to the on-chip buffers.

\setcounter{figure}{0}
\captionsetup[figure]{font=small, labelfont={bf},labelformat={default},labelsep=period,name={Figure }}
\begin{figure}[t] %%% t: top, b: bottom, h: here
\begin{center}
\includegraphics[width=1.0\linewidth]{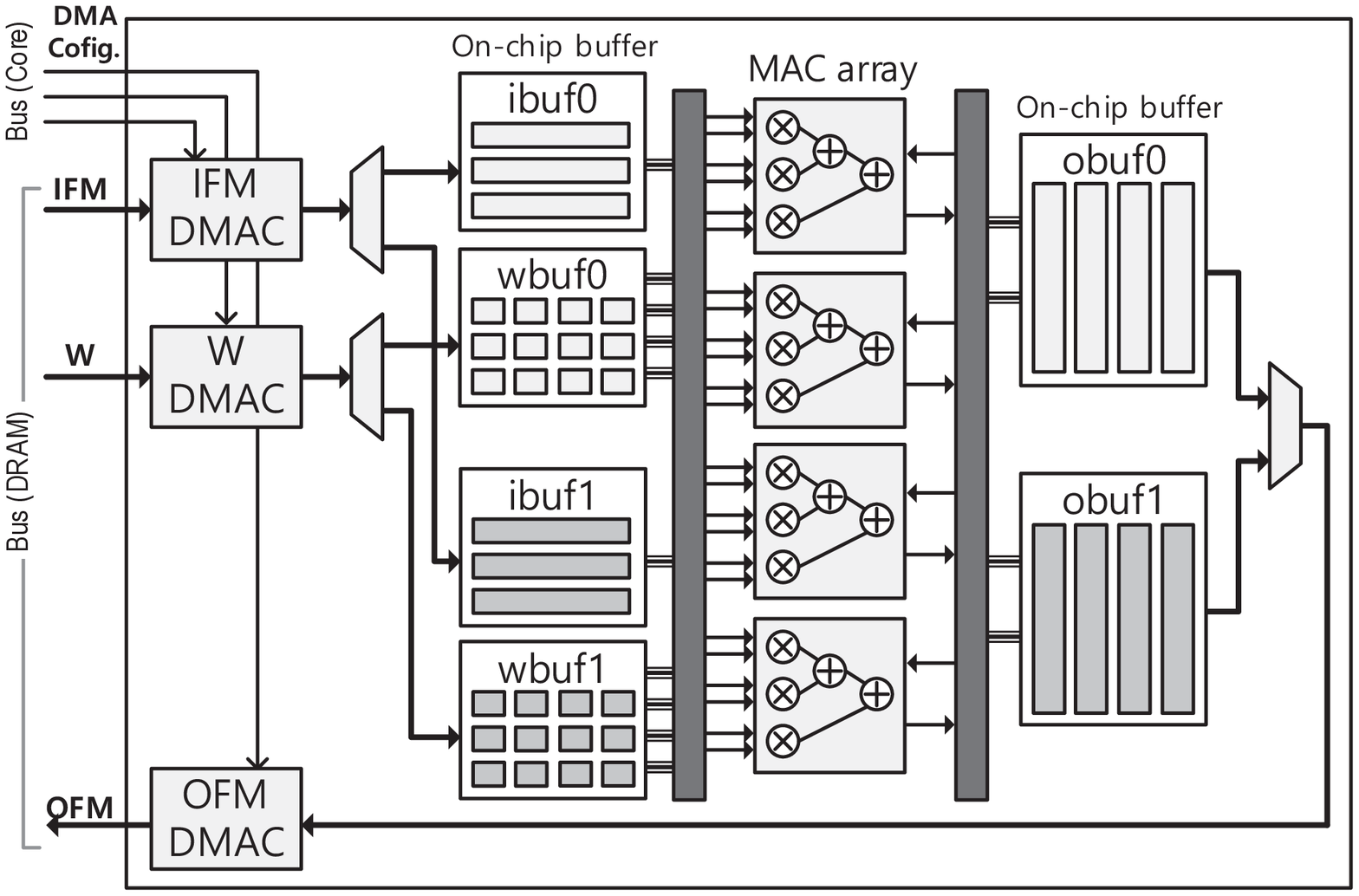}
\end{center}
%\captionof{lstlisting){-}
\caption{CNN accelerator for NLR with loop tiling}
\label{lst:1}
\end{figure}

\subsection{System under Consideration}

Figure 2 illustrates the system assumed in this paper, which consists of an accelerator, a processor core, a DRAM controller and system buses. The processor core is responsible for synchronizing DMACs to each processing pass. It also configures DMACs by setting the transfer addresses and sizes. The DRAM controller assists the accelerator in its access to off-chip DRAM. The system buses connect the aforementioned hardware blocks together. Since an accelerator is usually designed as a reusable IP block, a standardized interface may ease the integration into the system. In this work, we assume the AMBA AXI4 interface, in detail, AXI-MM for data transfer (DRAM controller) and AXI-Lite for command transfer (processor core).

Figure 3 exemplifies the operation of a convolutional layer with 12 input feature maps (C = 12) and 12 filters (M = 12). The tile sizes are set as TB = 1, TC = 3, TM = 4, TE = 5 and TF= 5. Assuming the output feature maps of 10 x 10 (i.e., E = F = 10), one convolutional layer consists of a total of 48 processing passes. It is clearly shown in the figure that the accelerator hardware is pipelined over consecutive processing passes. The figure also shows that the input feature maps and filters are loaded from DRAM every processing pass, whereas the output feature maps are stored into DRAM every 4 passes. This is consistent with the buffer switching illustrated in the figure. It should be noted that the transfer of output feature maps is spread over multiple processing passes to reduce the  communication bandwidth. It should also be noted that the duration of a processing pass is determined by the maximum between the communication time (i.e., the time required for load/store operations) and the computation time (i.e., the time required for MAC operations). Depending on which is longer, it is referred to as communication-limited or computation-limited [2], [3]. The figure illustrates a communication-limited case, since the communication time is longer.

\begin{figure}[t] %%% t: top, b: bottom, h: here
\begin{center}
\includegraphics[width=.8\linewidth]{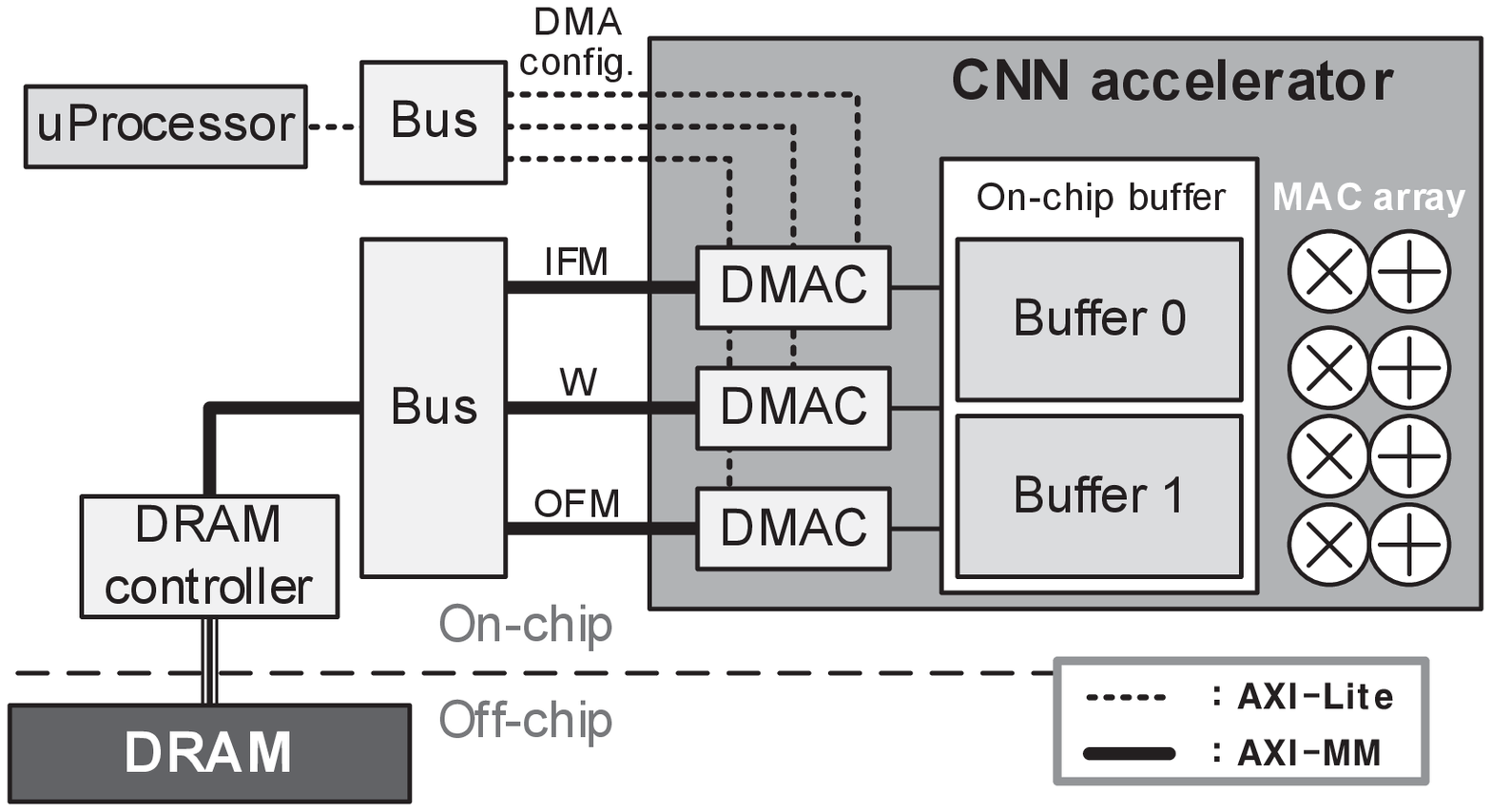}
\end{center}
%\captionof{lstlisting){-}
\caption{Overall system with a CNN accelerator as a reusable IP block}
\label{lst:1}
\end{figure}

\begin{figure}[t] %%% t: top, b: bottom, h: here
\begin{center}
\includegraphics[width=1.0\linewidth]{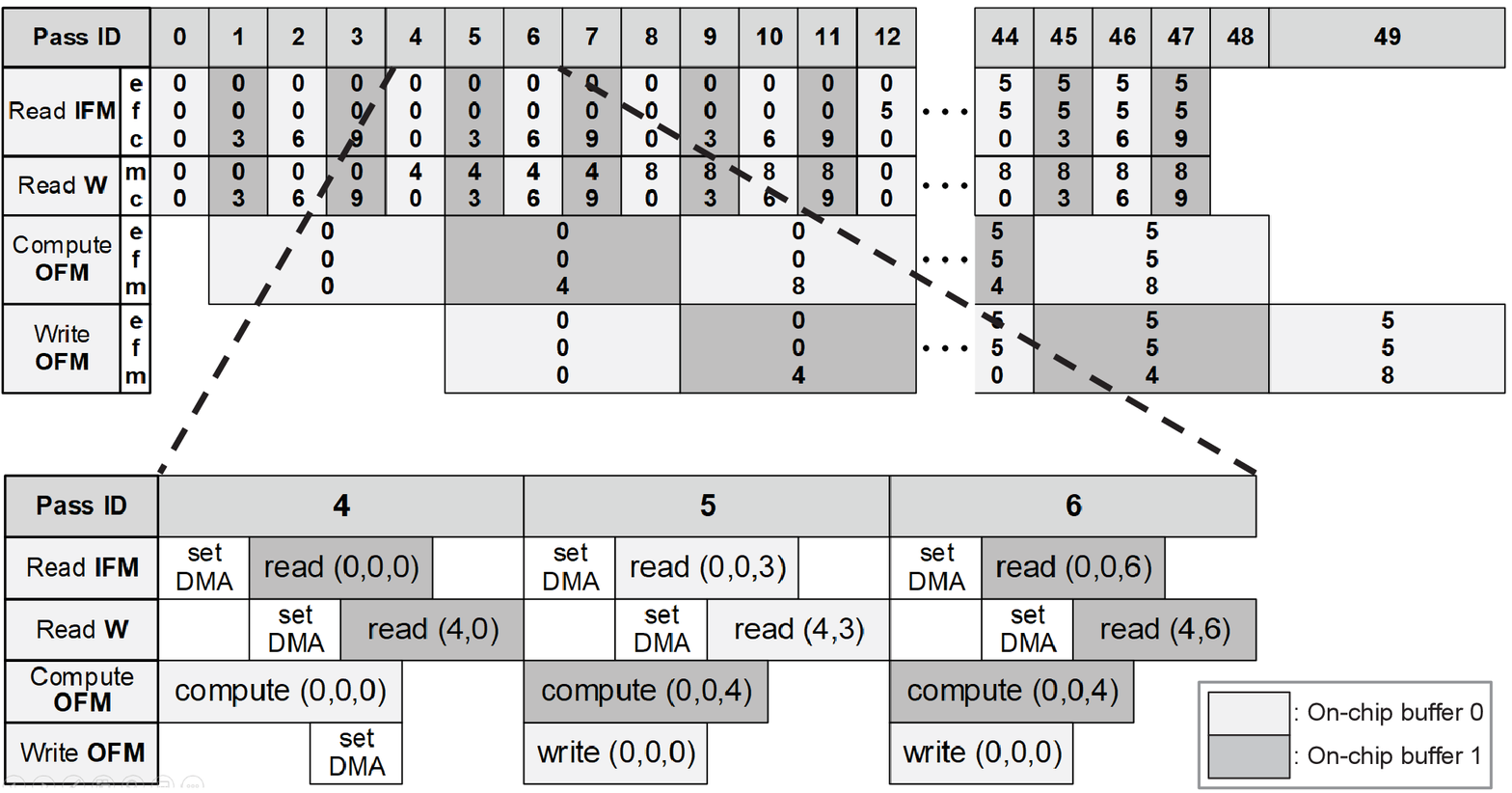}
\end{center}
%\captionof{lstlisting){-}
\caption{Processing passes in a convolutional layer}
\label{lst:1}
\end{figure}

\section{Accelerator Simulator}

This section describes the proposed AccTLMSim, a pre-RTL cycle-accurate accelerator simulator, which is based on SystemC TLM. The simulation model for the CNN accelerator, in particular, the DMACs, is depicted in detail, focusing on the cycle-accurate modeling on the model boundary.

\subsection{Transaction-level Modeling}

Figure 4 shows how the system described in Figure 2 is modeled in AccTLMSim. Each of the hardware blocks is implemented as a SystemC module (sc\_module) with sockets, except the DRAM subsystem taken from an open source simulator, DRAMSim2 [19]. As is typical in the TLM-based simulators [20]-[24], AccTLMSim provides the capability to skip the actual processing inside an individual module for simulation speedup (e.g., computation of the MAC array) while maintaining the cycle-accurate timing on the module boundary, i.e., on the socket level.

In order to model the AXI protocol efficiently, the GSGP sockets [25] as well as the TLM2.0 sockets are used. More specifically, the GSGP sockets are used for AXI-MM while the TLM2.0 sockets are for AXI-Lite. Figure 5 shows the generic protocol supported by the GSGP sockets where its six phases are mapped into the handshake signals (i.e., valid and ready) of the five AXI channels – AR, AW, R, W and B [26]. Compared to the TLM2.0 base protocol, the two phases, \texttt{BEG\_DAT} and \texttt{END\_DAT} are newly added to model the handshaking of the W channel. The AXI protocol supports a burst transaction and it is possible to keep track of every handshake of a burst transaction by iteratively using \texttt{BEG\_RSP} and \texttt{END\_RSP} for a read AXI burst, and \texttt{BEG\_DAT} and \texttt{END\_DAT} for a write AXI burst. Note that such a low-level modeling enables us to evaluate the impact of several protocol-related parameters such as the number of outstanding transactions.

\begin{figure}[t] %%% t: top, b: bottom, h: here
\begin{center}
\includegraphics[width=.75\linewidth]{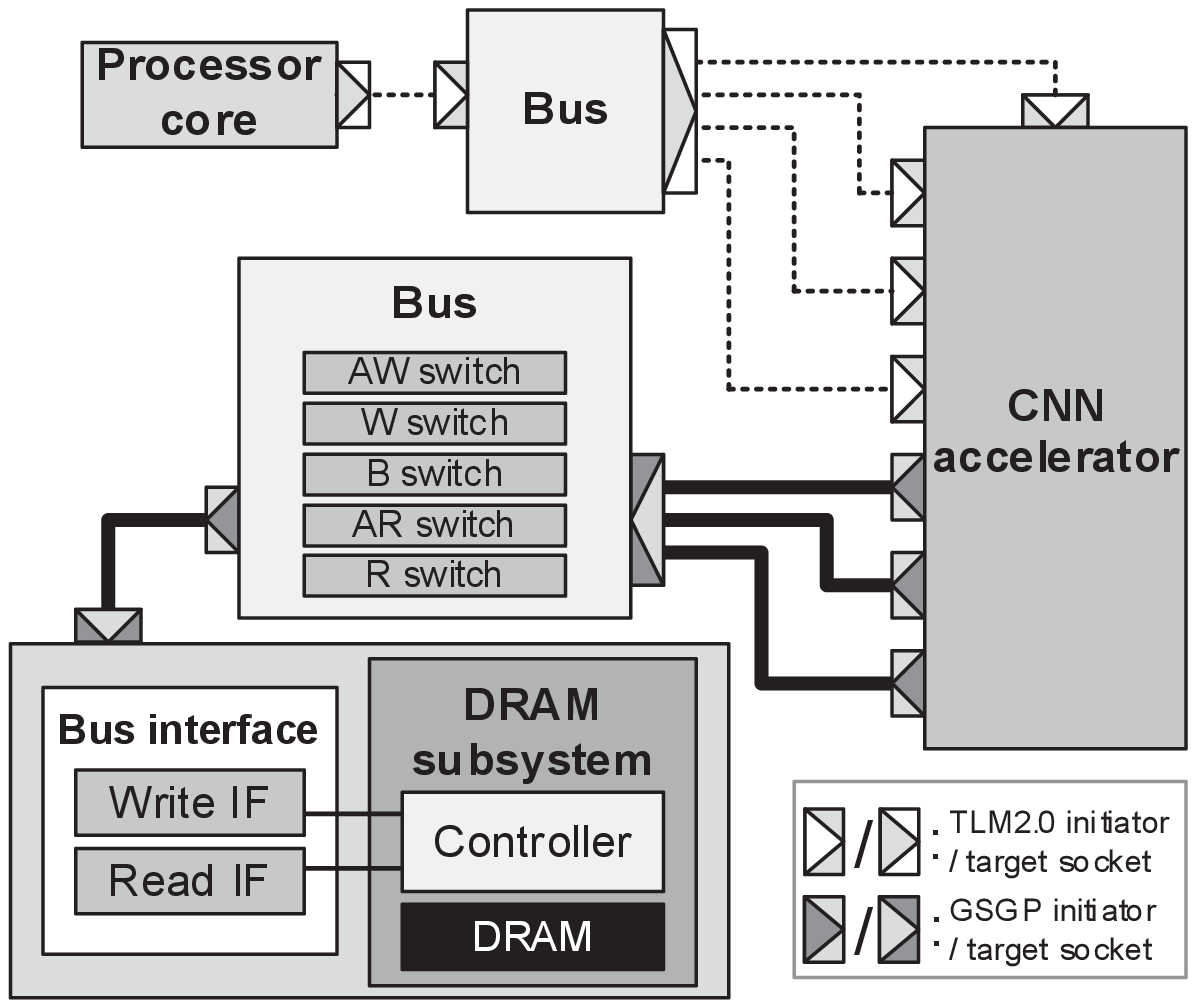}
\end{center}
%\captionof{lstlisting){-}
\caption{AccTLMSim: TLM based accelerator simulator for CNNs}
\label{lst:1}
\end{figure}

\begin{figure}[t] %%% t: top, b: bottom, h: here
\begin{center}
\includegraphics[width=.55\linewidth]{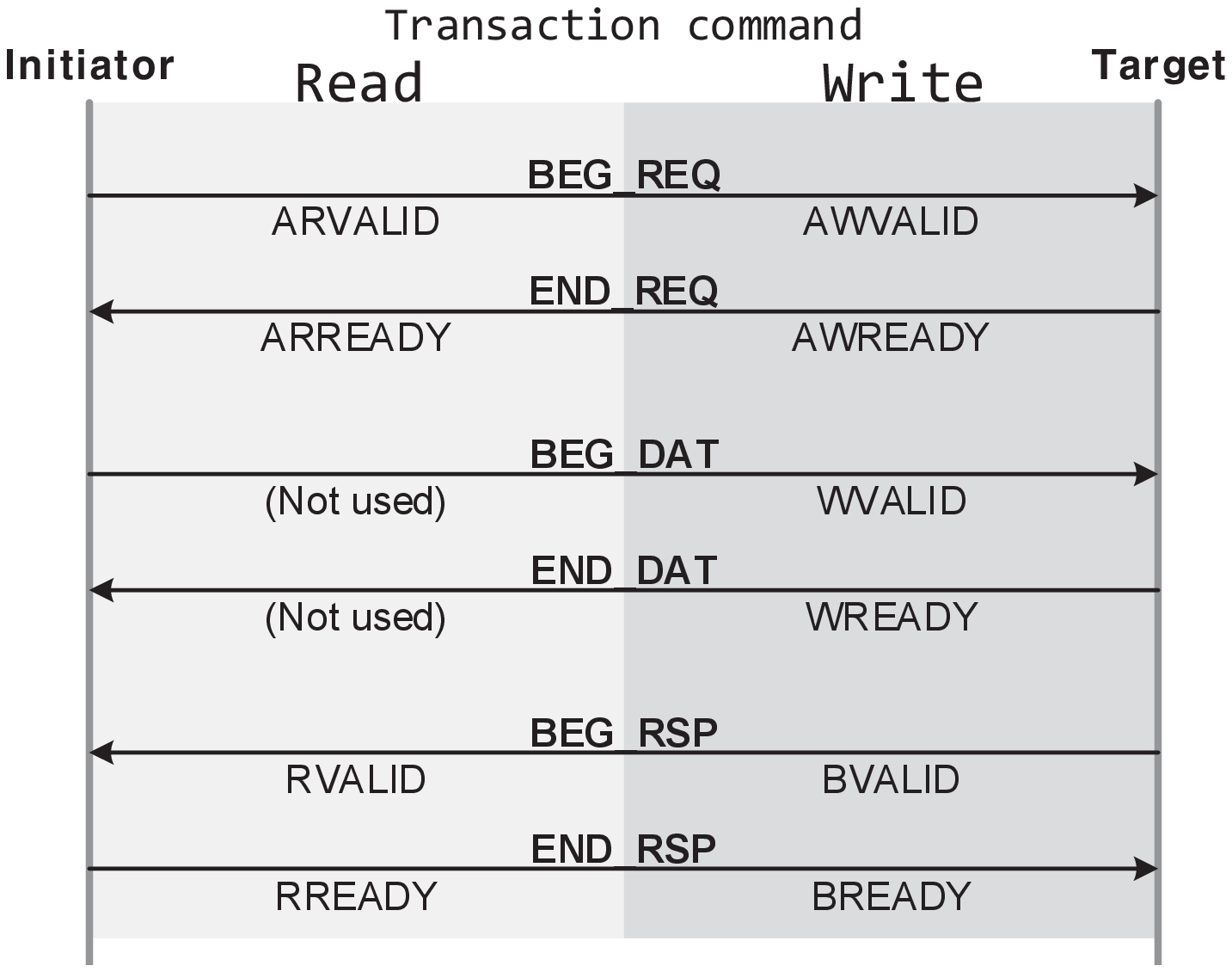}
\end{center}
%\captionof{lstlisting){-}
\caption{Generic protocol using the GSGP sockets for AXI-MM}
\label{lst:1}
\end{figure}

The sockets make use of the TLM2.0 core interfaces to communicate between hardware blocks. In general, the approximately-timed coding style based on non-blocking transports (nb\_transport) is chosen for data transfer whereas the loosely-timed coding style based on blocking transports (b\_transport) is chosen for control transfer [27]. It is also worth mentioning that every transfer between hardware blocks is modeled in an event-driven yet cycle-accurate manner. For example, it is possible to keep track of the data transfer between DMACs and on-chip buffers every cycle.

Figure 6 provides a detailed description of the simulation model that is commonly used in AccTLMSim. In the figure, an interconnect receives data from an initiator and relays to a target. These modules are connected together through the TLM2.0 socket. The handshaking between modules is modeled using the TLM2.0 core interfaces. This figure shows the approximately-timed coding style based on the non-blocking transport interfaces. The TLM2.0 payload (generic payload) is used to convey the relevant information such as the transfer address, size and phase. For example, the initiator sends a request through \texttt{interface0}, resulting in insertion into payload event queue (peq). After a predefined handshaking delay, a callback function (peq cb) is called to respond to the initiator (through an interface of the initiator). The callback function also notifies the interconnect thread, \texttt{thread0} (sc thread) of this handshaking through \texttt{event0} (sc event) and, at the same time, it pushes the received request to the internal FIFO (sc fifo). Subsequently, \texttt{function0} performs the pre-defined task, e.g., delayed arbitration, and the interconnect thread sends the received request (pulled from the internal FIFO) to the target (through an interface of the target, \texttt{thread1}). The target responds to this request through \texttt{interface1}, which again notifies the interconnect thread of this handshaking. A simplified version of the corresponding pseudo code is given in Listing 3. This description is generally applicable to several simulation models such as those for system buses and DMACs.

\begin{figure}[t] %%% t: top, b: bottom, h: here
\begin{center}
\includegraphics[width=.8\linewidth]{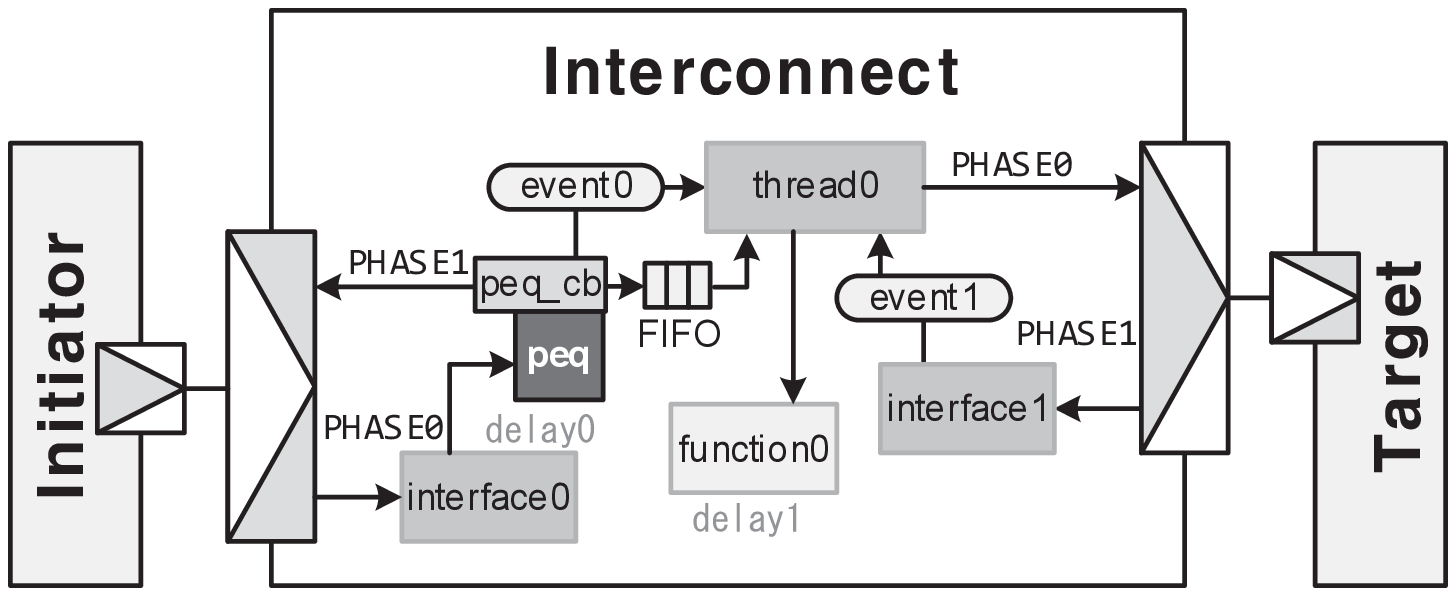}
\end{center}
%\captionof{lstlisting){-}
\caption{Simulation model that is commonly used in AccTLMSim}
\label{lst:1}
\end{figure}

\setcounter{figure}{2}
\captionsetup[figure]{font=small, labelfont={bf},labelformat={default},labelsep=period,name={Listing }}
\begin{figure}[t] %%% t: top, b: bottom, h: here
\begin{center}
\includegraphics[width=1.0\linewidth]{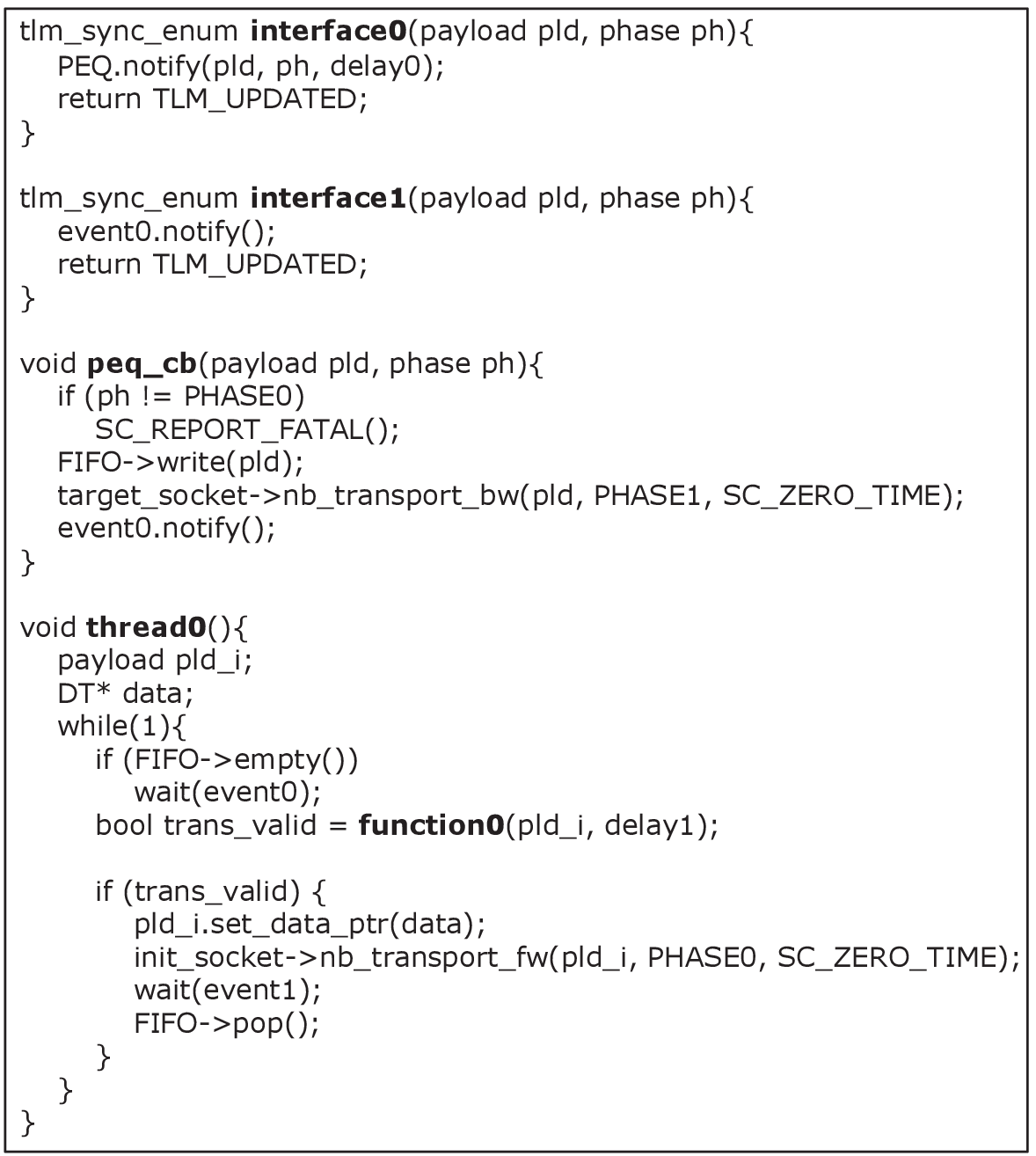}
\end{center}
%\captionof{lstlisting){-}
\caption{Pseudo code for the simulation model depicted in Figure 6}
\label{lst:1}
\end{figure}
\setcounter{figure}{0}

\subsection{Accelerator Modeling}

As shown in Figure 7, the accelerator module is comprised of several submodules – DMACs, stream interfaces, NoC, MAC array and on-chip SRAM. The stream interfaces are the submodules that connect the DMACs to the on-chip SRAM. In detail, they generate the address and control signals required by the on-chip SRAM. The NoC is responsible for moving data between the on-chip SRAM and the MAC array. The specific architecture is implement-dependent, for example, a crossbar bus [2], a set of shared buses [8] or a hierarchical mesh [9]. The accelerator controller is modeled as a thread, which exchanges a set of control signals with the aforementioned submodules, for example, with the stream interface to enable double buffering. The corresponding control transfer is implemented using the loosely-timed coding style based on blocking transport, as mentioned before. It is also shown that the TLM2.0 sockets are used to connect the submodules whereas the GSGP sockets are used to connect the DMACs to the external of the accelerator.

Figure 8 describes the simulation model for DMACs in more detail. As shown in the figure, a DMAC receives a DMA command from a processor core and, according to the command, it reads data from or writes to the DRAM subsystem. It is connected to either the processor core or the stream interface through the TLM2.0 sockets while it is connected to the DRAM through the GSGP sockets. The DMA command includes the burst addresses and lengths of the bus requests and it is passed to an AXI thread through the command FIFO. On the bus side, the AXI thread sends a request to the DRAM subsystem (i.e., asserting ARVALID or AWVALID) and then the DRAM subsystem responds to this request through an interface (i.e., asserting ARREADY or AWREADY). The handshakings of the remaining AXI channels are modeled in a similar manner. When a burst transaction is completed, the AXI thread is notified of the event and then it begins to send the next request to the DRAM. On the stream interface side, the DMAC sends a request or responds to the stream interface using the stream thread. Taking into account the stream direction, the IFM/W DMAC is modeled as a master whereas the OFM DMAC is modeled as a slave. This explains the definition of a backward transport in the IFM/W DMAC and a forward transport in the OFM DMAC. The AXI protocol supports  multiple outstanding transactions [26] and thus the simulation model is parameterized accordingly, for example, by the maximum number of outstanding bus requests and the latency between consecutive outstanding bus requests. After completing the DMA command sent by the processor core, the status register is updated accordinlgy so that the processor can figure out whether the DMAC has completed the corresponding transaction.

\setcounter{figure}{6}
\captionsetup[figure]{font=small, labelfont={bf},labelformat={default},labelsep=period,name={Figure }}
\begin{figure}[t] %%% t: top, b: bottom, h: here
\begin{center}
\includegraphics[width=1.0\linewidth]{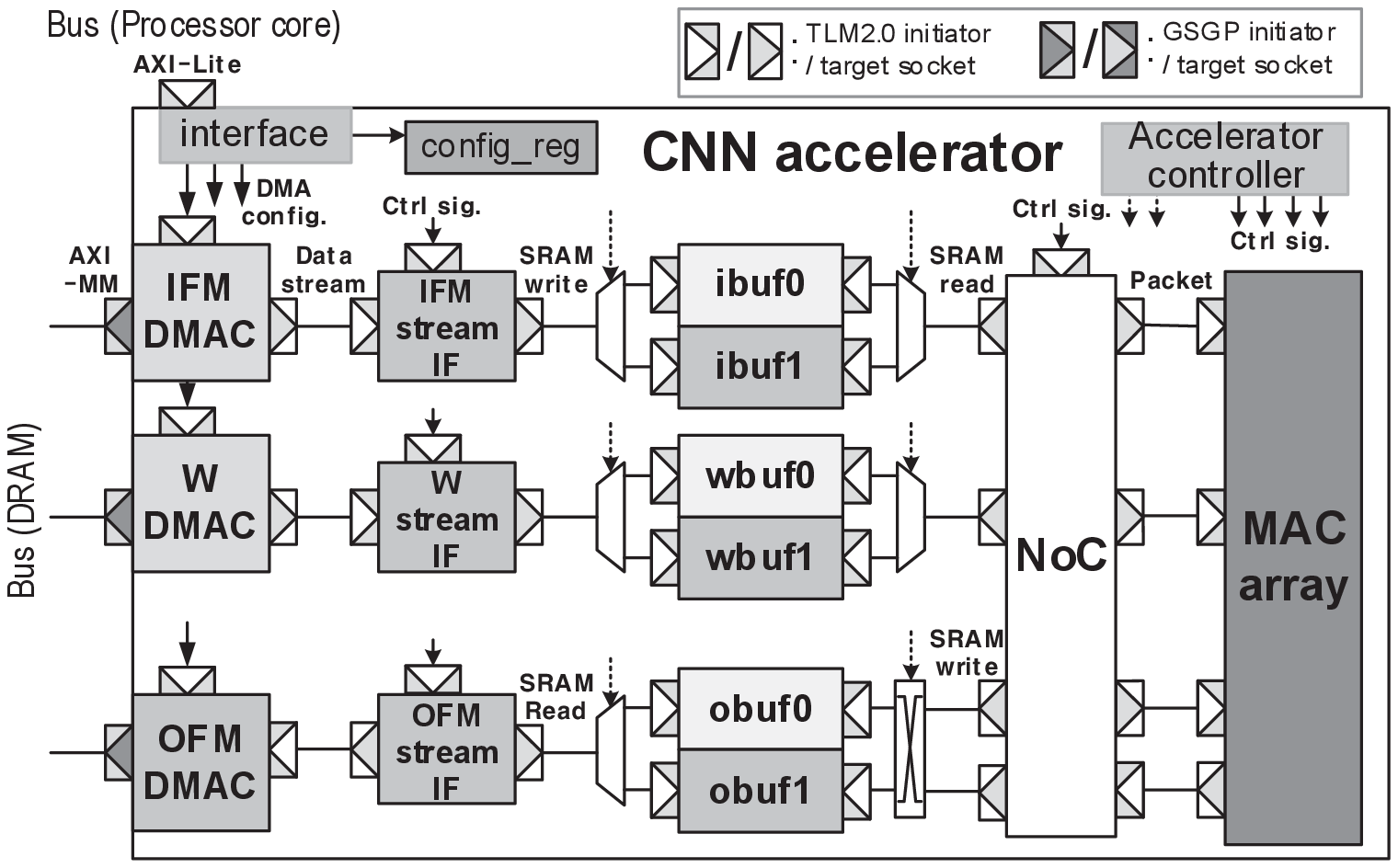}
\end{center}
%\captionof{lstlisting){-}
\caption{Simulation model for CNN accelerator}
\label{lst:1}
\end{figure}

\begin{figure}[t] %%% t: top, b: bottom, h: here
\begin{center}
\includegraphics[width=.8\linewidth]{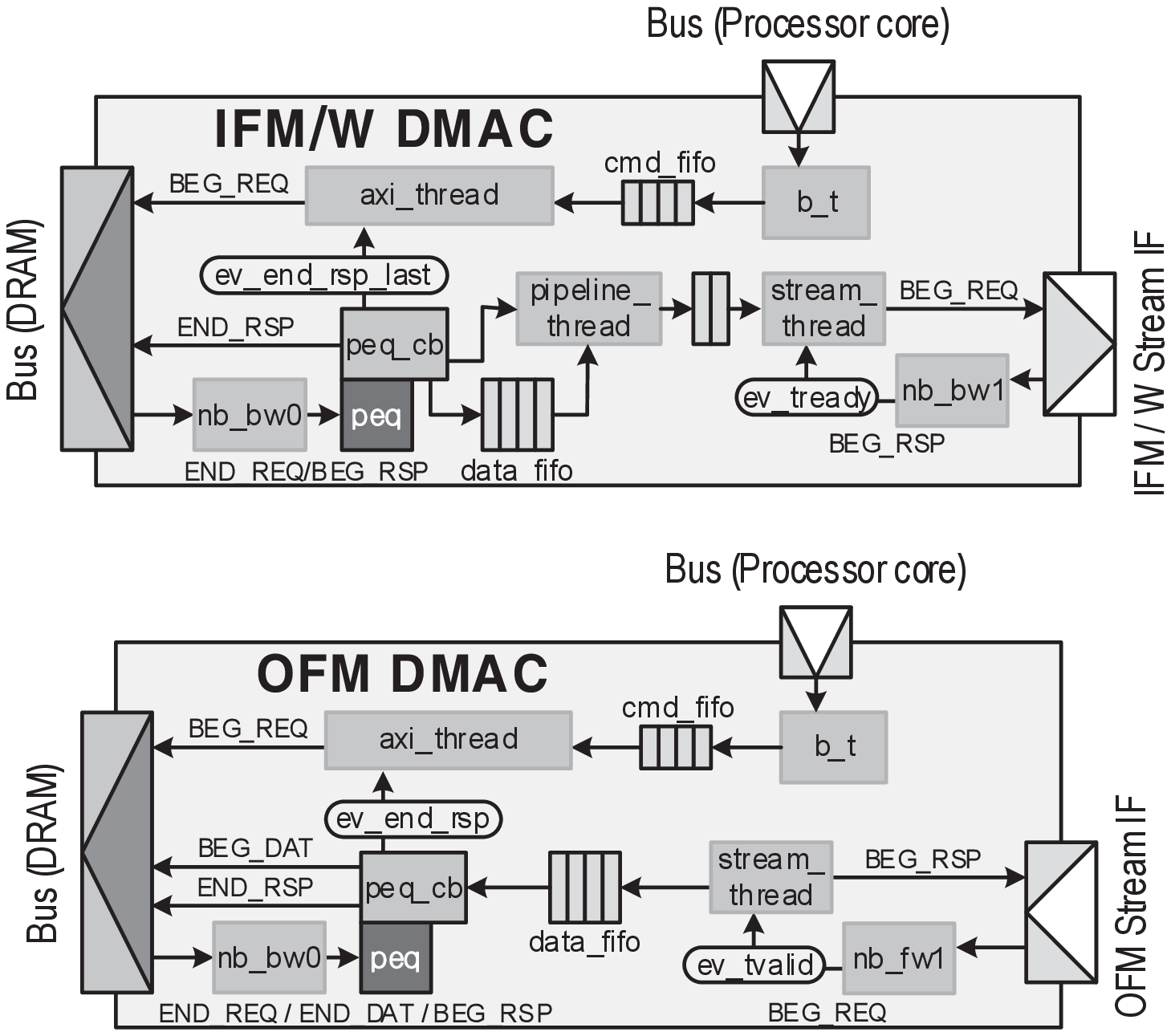}
\end{center}
%\captionof{lstlisting){-}
\caption{Simulation model for IFM/W DMAC (top) and OFM DMAC (bottom)}
\label{lst:1}
\end{figure}

\subsection{Validation against Implementations}

The simulation results of AccTLMSim were validated against the implementation results. First, the simulation models for the CNN accelerator and the system buses were validated against the register transfer level (RTL) implementations. Here the accelerator was implemented using the Vivado High-Level Synthesis (HLS) [28] and was further customized, for example, the bus interface including the scatter-gather DMACs. It was verified that the data transfer between the accelerator and the DRAM subsystem is modeled cycle-accurately. Second, the rest of the AccTLMSim, the DRAM subsystem taken from the DRAMSim2, was validated against the Xilinx Zynq 7000 (ZC7045) on the ZC706 evaluation board [29]. In detail, the implemented DRAM subsystem consists of Xilinx Memory Interface Generator and Micron DDR3 SDRAM (MT8JTF12864). The simulation model for the DRAM subsystem was empirically tuned to mitigate the mismatch due to the relevant implementation ambiguity.

Figure 9 compares the simulation results of AccTLMSim with the implementation results. In addition, some of the simulation models are replaced by those from Platform Architect [16] (DRAM subsystem and system buses), DRAMsim2 [19] (DRAM subsystem) and Ramulator [30] (DRAM subsystem). After taking a closer look at the simulation and implementation results, it was found out that the major source of modeling error lies in the difference of the DRAM controller modes and policies.

\begin{figure}[t] %%% t: top, b: bottom, h: here
\begin{center}
\includegraphics[width=1.0\linewidth]{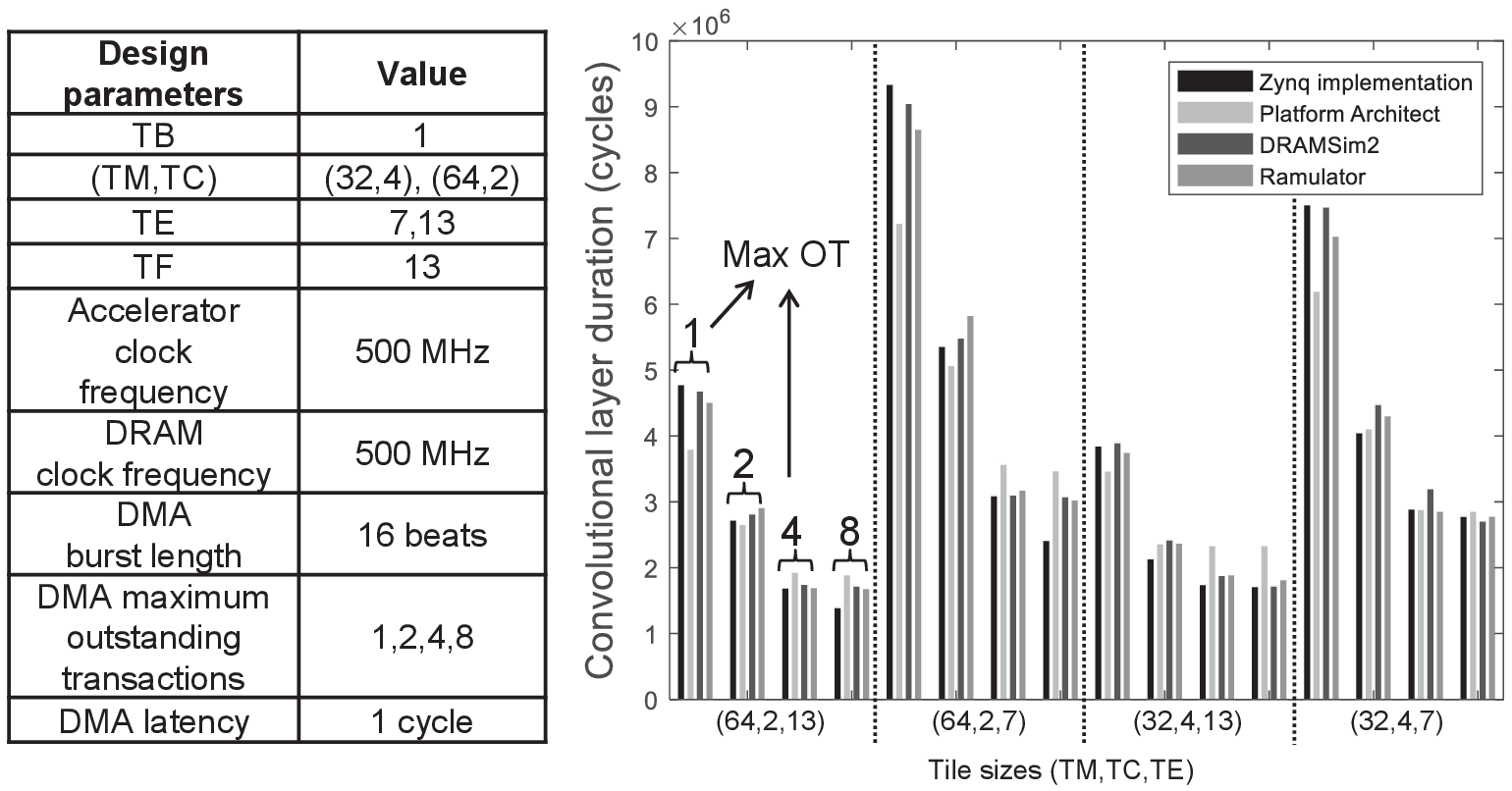}
\end{center}
%\captionof{lstlisting){-}
\caption{Simulator validation with different simulation models}
\label{lst:1}
\end{figure}

\subsection{Simulation Speed}

The proposed accelerator simulator is based on the event-driven modeling except for the DRAM subsystem that is taken from the DRAMSim2, a cycle-driven simulator [19]. Setting the ease of modeling apart, the SystemC TLM simulations are generally orders of magnitude faster than the RTL simulations [31]. For the design parameters given in Table I, the average simulation time was roughly 2.1 seconds per convolutional layer (per image).

\begin{table}[t]
  \centering
  \caption{Design parameters for measuring simulation speed }
  \begin{tabular}{|c|c|c|}
    \hline
    \centering
    \textbf{Design parameters} & \textbf{Value}\\
    \hline
    \hline
    \centering
    TB & 1:3 \\
    \hline
    \centering
    (TM,TC) & (21,6), (42,3), (64,2)\\
    \hline
    \centering
    TE & 7, 13\\
    \hline
    \centering
    TF & 13\\
    \hline
    \centering
    Accelerator clock frequency & 500 MHz\\
    \hline
    \centering
    DMA burst length                                                                                    & 16 beats\\ \hline
    \centering
    \begin{tabular}[c]{@{}c@{}} DMA maximum\\ outstanding transactions\end{tabular}         & 2 \\ \hline
    \hline
    \centering
    DMA latency & 5 cycles \\
    \hline
  \end{tabular}
  \label{table:formatting}
\end{table}

\section{Performance Estimation}

In this section, the proposed performance estimation model is described. Before going into the details, the impact of tile sizes on the accelerator performance is explained.

\subsection{Performance Impact of Tile Sizes}

The communication bandwidth, which is defined as the number of pixels per cycle, is the most important factor to determine the performance. Note that the communication bandwidth tends to be lower than unity, and, more importantly, it is often dynamic, varying with time. Figure 10 shows the communication bandwidth measured by the aforementioned accelerator simulator, assuming the 3rd convolutional layer of AlexNet with R = 3, S = 3, E = 13 and F = 13 [32]. Here the tile sizes are set as TB = 1, TC = 2, TM = 64, TE = 6 and TF= 13. In the figure, depending on the set of active DMACs, a processing pass is divided into five intervals, each of which is referred to as a DMA interval. For example, the OFM DMAC runs in the first DMA interval, both the IFM DMAC and the OFM DMAC run in the second DMA interval, all three DMACs run in the third DMA interval and so on. The duration of a DMAC interval depends when a DMAC starts running or stops running. Figure 10 (a) shows that the communication bandwidth varies across DMA intervals.

As pointed out earlier, it is either DRAM latency or bus protocol overhead that limits the communication bandwidth. Let us take a closer look at the waveform views generated by the visualization tool provided by AccTLMSim – Figure 10 (b) and (c). In case that the bus requests from the accelerator arrive at the DRAM controller at a sufficiently high rate, a bank of DRAM is fully populated with the corresponding DRAM commands (i.e., activate, read, write or precharge), in other words, the communication bandwidth is limited by DRAM latency, as shown in Figure 10 (b). Assuming that a DMAC support up to two outstanding bus requests, it is shown that a set of outstanding bus requests leads to a single page open followed by a series of DRAM commands – activate first and precharge last. Note that the bus requests from one DMAC may cause DRAM contention with those from other DMACs, making each bus request experience a dynamic DRAM latency. In this case, the communication time of accelerator can be calculated by the DRAM execution time alone, as shown in Figure 10 (b). On the other hand, in case that the bus requests arrive at a slower rate, the bank remains idle waiting for the next bus request for most of the time, as shown in Figure 10 (c). In other words, the communication bandwidth is limited by bus protocol overhead. It is shown that, in this case, the communication time of accelerator can be calculated by summing the DRAM/bus latency as well as the DRAM execution time, as opposed to the aforementioned DRAM-limited case. Thus the communication bandwidth generally decreases with the DRAM execution time, whether it is limited by DRAM latency or bus protocol overhead.

\begin{figure}	
	\begin{subfigure}{.5\textwidth}
		\centering
		\includegraphics[width=2.5in]{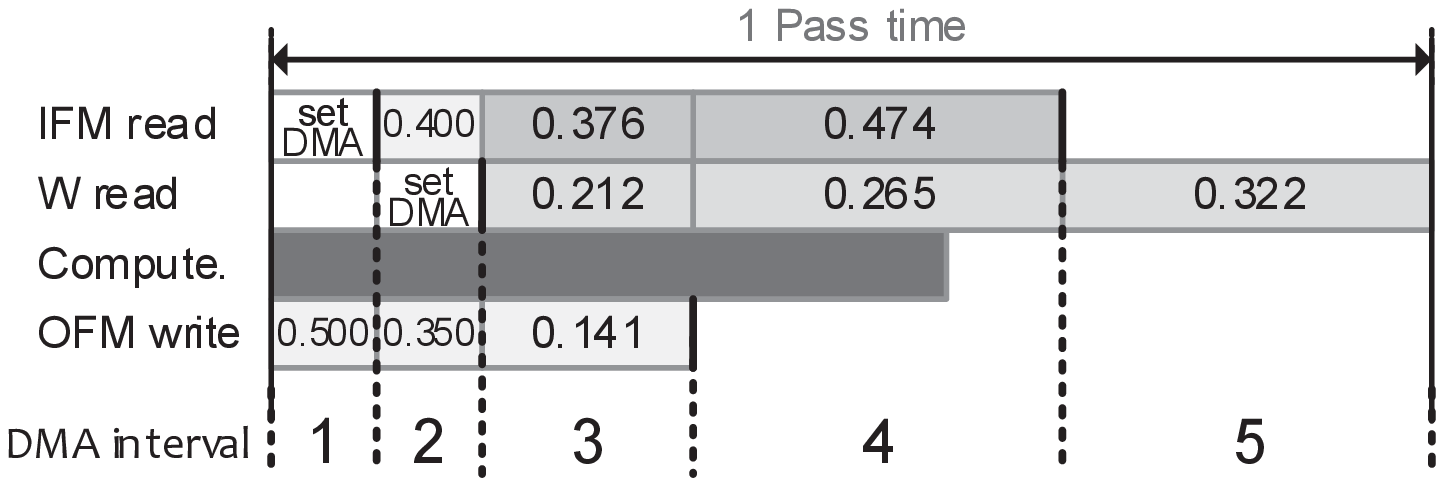}
		\caption{}\label{fig:1a}		
	\end{subfigure}
    \newline
	\begin{subfigure}{.5\textwidth}
		\centering
		\includegraphics[width=3.5in]{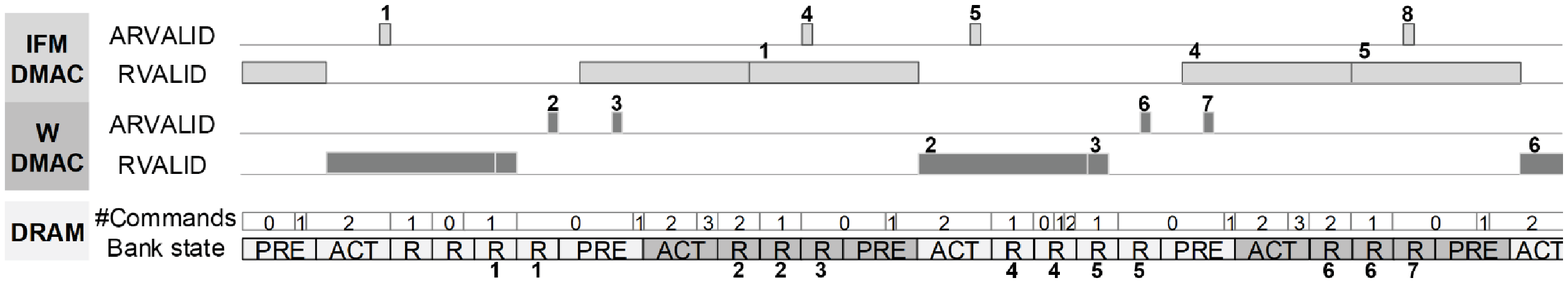}
		\caption{}\label{fig:1a}		
	\end{subfigure}
	\newline
	\begin{subfigure}{.5\textwidth}
		\centering
		\includegraphics[width=3.5in]{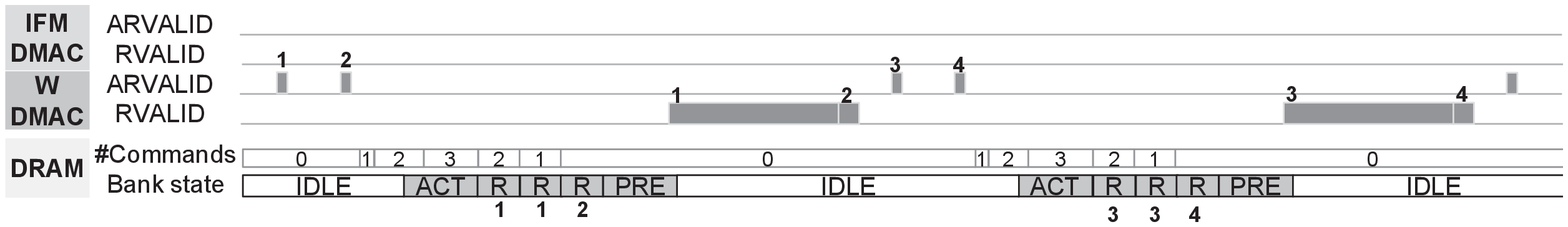}
		\caption{}\label{fig:1a}		
	\end{subfigure}
	\caption{Impact of DRAM latency and bus protocol overhead on communication bandwidth for
    (a) communication bandwidth in a processing pass,
    (b) communication bandwidth limited by DRAM latency and
    (c) communication bandwidth limited by bus protocol overhead}
    \label{fig:1}
\end{figure}

The DRAM execution time increases with the total number of DRAM commands, which is determined by the burst addresses and lengths of the outstanding bus requests. In general, the more contiguous the burst addresses are, the fewer page opens the outstanding bus requests result [33]. In particular, the outstanding bus requests may result in even a single page open if they arrive before the page closes. In addition, it is possible to reduce the number of page opens further by increasing the burst lengths even when the outstanding bus requests arrive at a low rate. Recall that, for a given amount of communication data, the total number of DRAM commands increases with the number of page opens because of additional activates and precharges. Therefore, we can draw a conclusion that it is possible to reduce the DRAM execution time by either making the burst addresses more contiguous or increasing the burst lengths. For example, in Figure 10 (b), the burst addresses from the same DMAC are on the same page, and, assuming the maximum bus burst length of 16, the burst lengths are set to (16, 16) and (16, 2) for input feature maps and filters, respectively (as will be explained in more detail later in Figure 12). Therefore, assuming the maximum DRAM burst length of 8, a set of outstanding bus requests leads to a single page open with 6 DRAM commands (4 reads) and 5 DRAM commands (3 reads) for input feature maps and filters, respectively, as shown in the figure.

In general, the burst addresses and lengths of bus requests are determined by the DRAM data layout [34]-[36] and access pattern [33]-[35]. The data layout assumed in this paper assigns the pixels of a feature map to the DRAM locations in a row-major order, followed by those of the next feature map. Likewise the weights of a filter plane are assigned in a row-major order, followed by those of the next filter plane, i.e., the next input channel first and then the next output channel. The access pattern assumed in this paper follows the same principle, but it is confined to a single tile since it is defined per processing pass. Figure 11 exemplifies the data layout and access pattern for filters. For illustration purpose, it is assumed that each row of the DRAM consists of only 9 columns. It is readily found in the figure that, when the loop tiling is used, it is the tile sizes that determine the burst addresses and lengths of bus requests. Letting a set of the contiguous DRAM locations be denoted by a contiguous dataset, each processing pass has a total of TM contiguous datasets, each of which contains R$\times$S$\times$TC weights. In Figure 11, the loop tiling with TC = 6, TM = 2 leads to 2 contiguous datasets, each consisting of 54 weights, while the others have smaller contiguous datasets. Note that such tile sizes with larger contiguous datasets result in fewer page opens and thus smaller DRAM execution time since the burst addresses are more contiguous and the burst lengths are larger.

\begin{figure}[t] %%% t: top, b: bottom, h: here
\begin{center}
\includegraphics[width=1.0\linewidth]{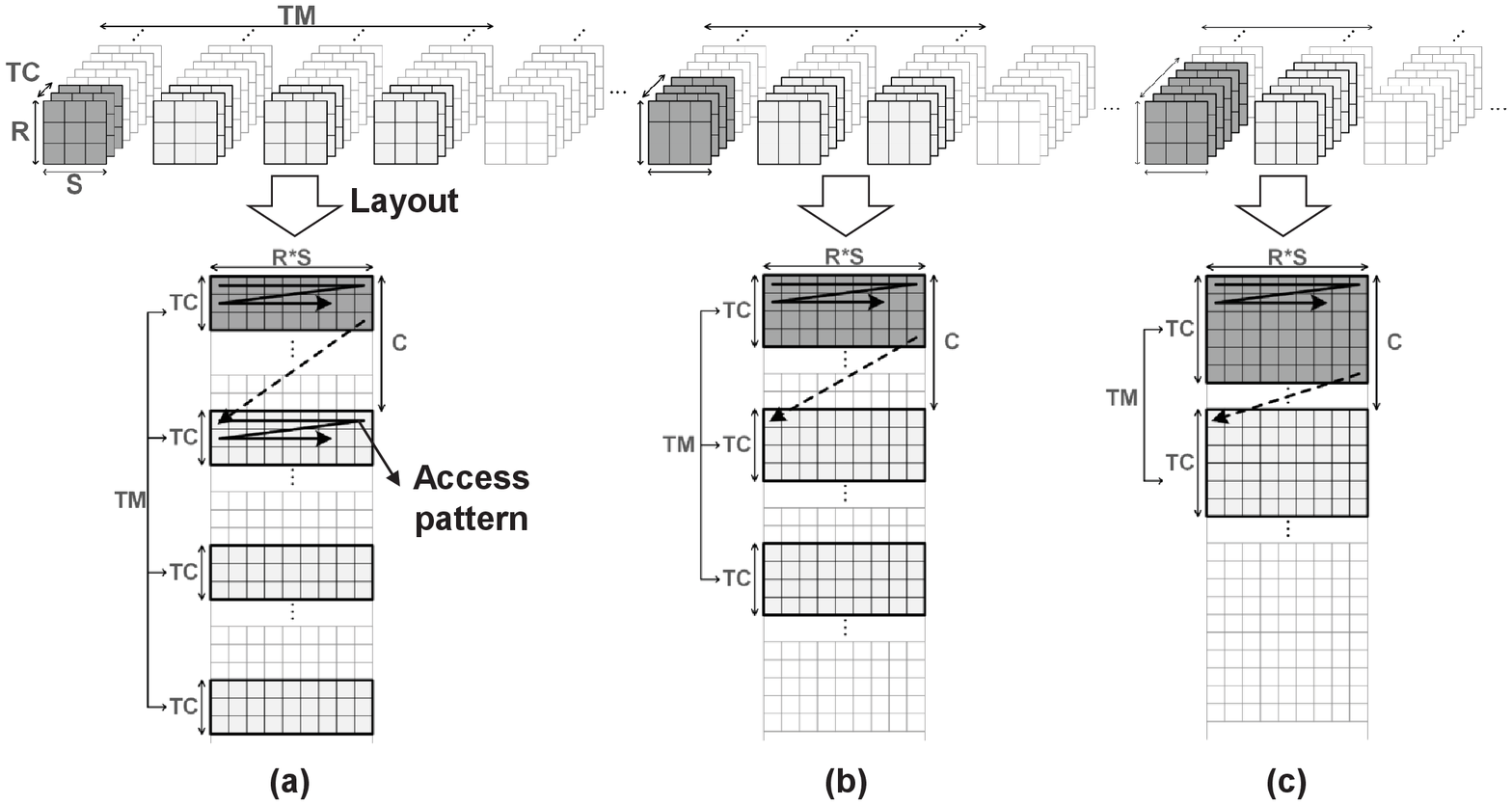}
\end{center}
%\captionof{lstlisting){-}
\caption{Impact of tile size on DRAM data layout and access pattern for
    (a) (TC=3, TM=4), (b) (TC=4, TM=3) and (c) (TC=6, TM=2)}
\label{lst:1}
\end{figure}

Figure 12 shows how a contiguous dataset is mapped into DMA bursts and DRAM commands assuming the same tile sizes as in Figure 10. It is shown that each contiguous dataset is accessed by one or more DMA bursts and executed by DRAM commands. For example, in the case of input feature map, a contiguous dataset consisting of 120 pixels is mapped into 8 DMA bursts and then 23 DRAM commands. Here we assume the maximum bus burst length of 16 and the maximum DRAM burst length of 8. Moreover, the open page mode with the 4-time close page policy is assumed [19], [24]. Thus it follows that the two consecutive DMA bursts result in a single page open with 5 or 6 DRAM commands (3 or 4 reads, respectively), as shown in the figure.

\begin{figure}[t] %%% t: top, b: bottom, h: here
\begin{center}
\includegraphics[width=1.0\linewidth]{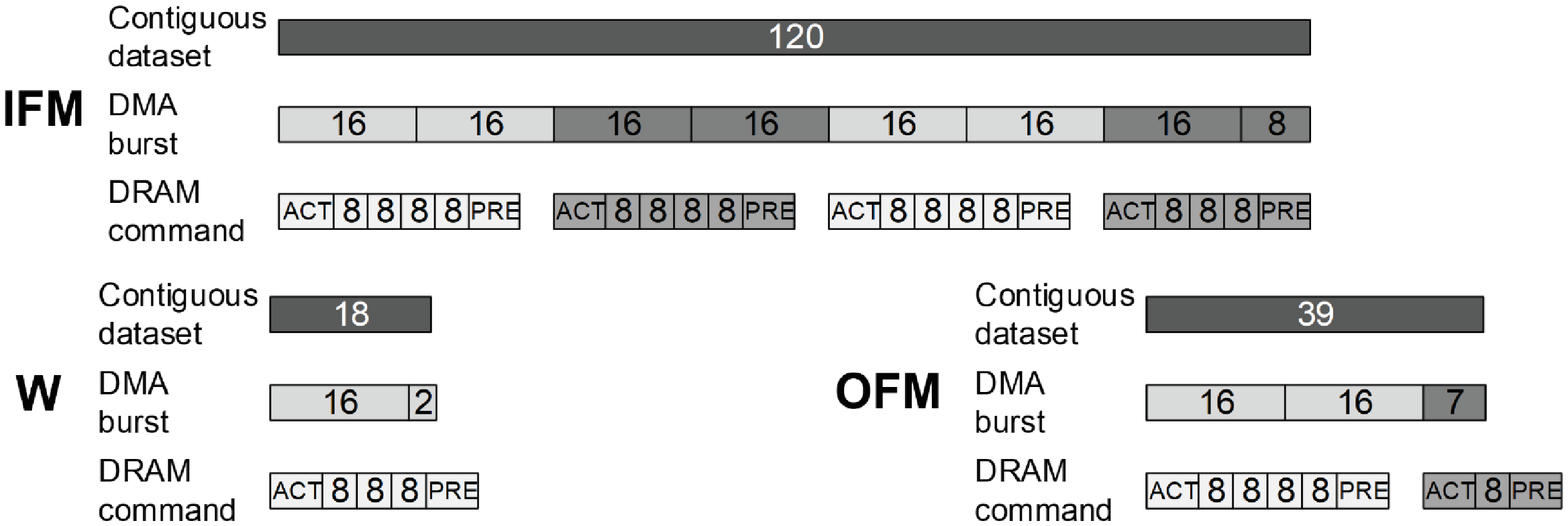}
\end{center}
%\captionof{lstlisting){-}
\caption{Mapping of a contiguous dataset into DMA bursts and DRAM commands}
\label{lst:1}
\end{figure}

\begin{figure}[t] %%% t: top, b: bottom, h: here
\begin{center}
\includegraphics[width=.95\linewidth]{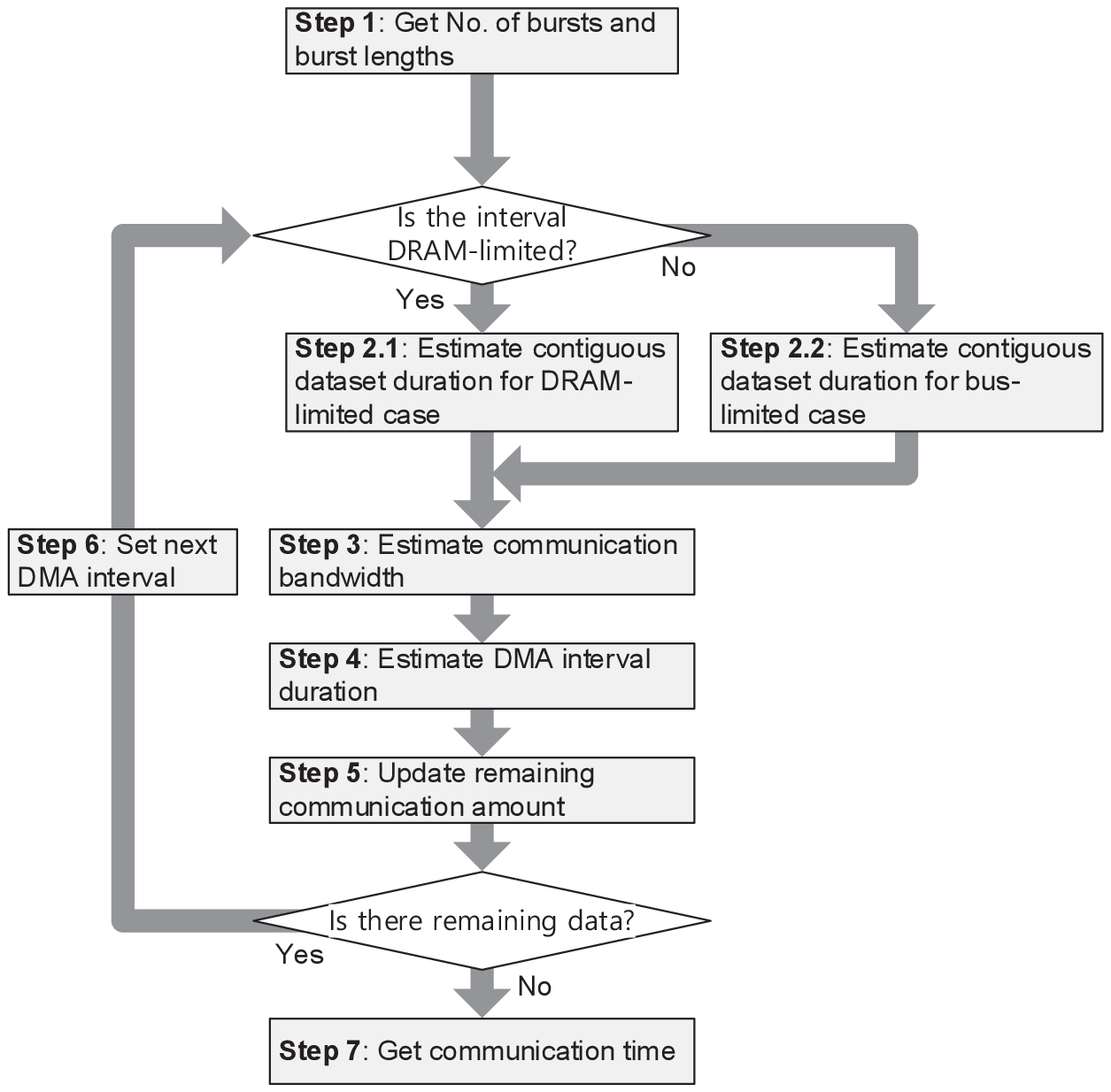}
\end{center}
%\captionof{lstlisting){-}
\caption{Proposed performance estimation algorithm}
\label{lst:1}
\end{figure}

\subsection{Proposed Performance Estimation Model}

The conventional model assumes an ideal communication bandwidth of one pixel per cycle [2],[4] and thus the communication time is assumed to be proportional to the communication amount, i.e., the number of pixels or weights. However, the communication bandwidth tends to be lower than unity and, more importantly, dynamic, as previously depicted in Figure 10. As a result, the communication time is not simply proportional to the communication amount, regardless of whether it is DRAM-limited or bus-limited. Thus we propose a new model to estimate the performance more accurately.

Figure 13 summarizes the proposed performance estimation model. As mentioned earlier, each processing pass is divided into several DMA intervals, depending on the set of active DMACs. Given the tile sizes, the proposed model first calculates the burst lengths of a contiguous dataset for each of the data types (Step 1). Once it figures out whether the DMA interval is limited by DRAM latency or bus protocol overhead, it estimates the duration of a contiguous dataset (Step 2). Subsequently, it estimates the communication bandwidth (Step 3) and then the duration of the DMA interval (Step 4). Once the remaining communication amount is updated (Step 5), it proceeds to the next DMA interval and repeats the aforementioned steps.

Listing 4 illustrates how to calculate the duration of a DMA interval. The parameters of the function include the communication bandwidth (\textsf{U}), the remaining communication amount (\textsf{P}), the size of a contiguous dataset (\textsf{C}) and the number of bursts per contiguous dataset (\textsf{Bc}). All the parameters are of the same size, which is given as the number of active DMACs (\textsf{D}). It first calculates the remaining communication amounts of the active DMACs (line 1). Here they are calculated in the unit of burst (i.e., not in the unit of pixels or weights) since all the active DMACs are assumed to have equal bus priority and send the same number of DMA bursts within the DMA interval. Subsequently, by taking the minimum of the remaining communication amounts (line 2), it calculates the actual communication amount of each active DMAC within the DMA interval (line 3). Finally, it calculates the duration of the DMA interval using the communication bandwidth (line 4) and updates the remaining communication amount for the next DMA interval (line 6).

\setcounter{figure}{3}
\captionsetup[figure]{font=small, labelfont={bf},labelformat={default},labelsep=period,name={Listing }}
\begin{figure}[t] %%% t: top, b: bottom, h: here
\begin{center}
\includegraphics[width=.55\linewidth]{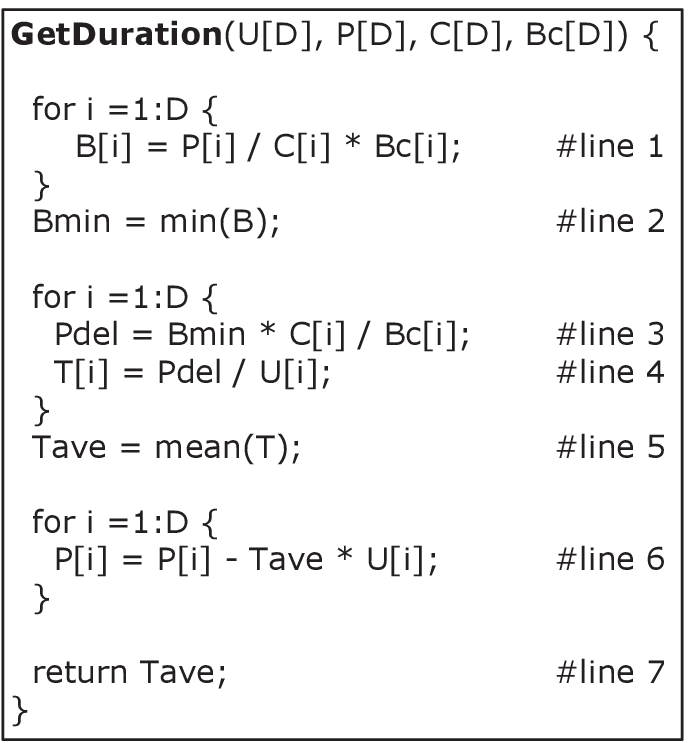}
\end{center}
%\captionof{lstlisting){-}
\caption{Pseudo code for estimating DMA interval duration}
\label{lst:1}
\end{figure}
\setcounter{figure}{0}

\section{Simulation Results}

Figure 14 presents the design space considered in this section. Assuming the 3rd convolutional layer of AlexNet, we parameterize the loop tiling by a predefined set of tile sizes given in the figure. The tile sizes for input channel (TC) and output channel (TM) are constrained to the maximum number of MAC units. It is also pointed out that each of the DMACs is assumed to support up to two outstanding bus requests with the burst length of 16. We assume the DRAM data layout and access pattern depicted in Section IV. When it comes to the DRAM controller, the scheduling is assumed to be the so-called first-ready, first-come-first-serve (FR FCFS) with the open page mode together with the 4-time close [19], [24], [33]. Finally, it is assumed that the accelerator operates at the same clock frequency as the off-chip DRAM. The design space shown in the figure consists of a total of 49,140 design points.

\setcounter{figure}{13}
\captionsetup[figure]{font=small, labelfont={bf},labelformat={default},labelsep=period,name={Figure }}
\begin{figure}[t] %%% t: top, b: bottom, h: here
\begin{center}
\includegraphics[width=.8\linewidth]{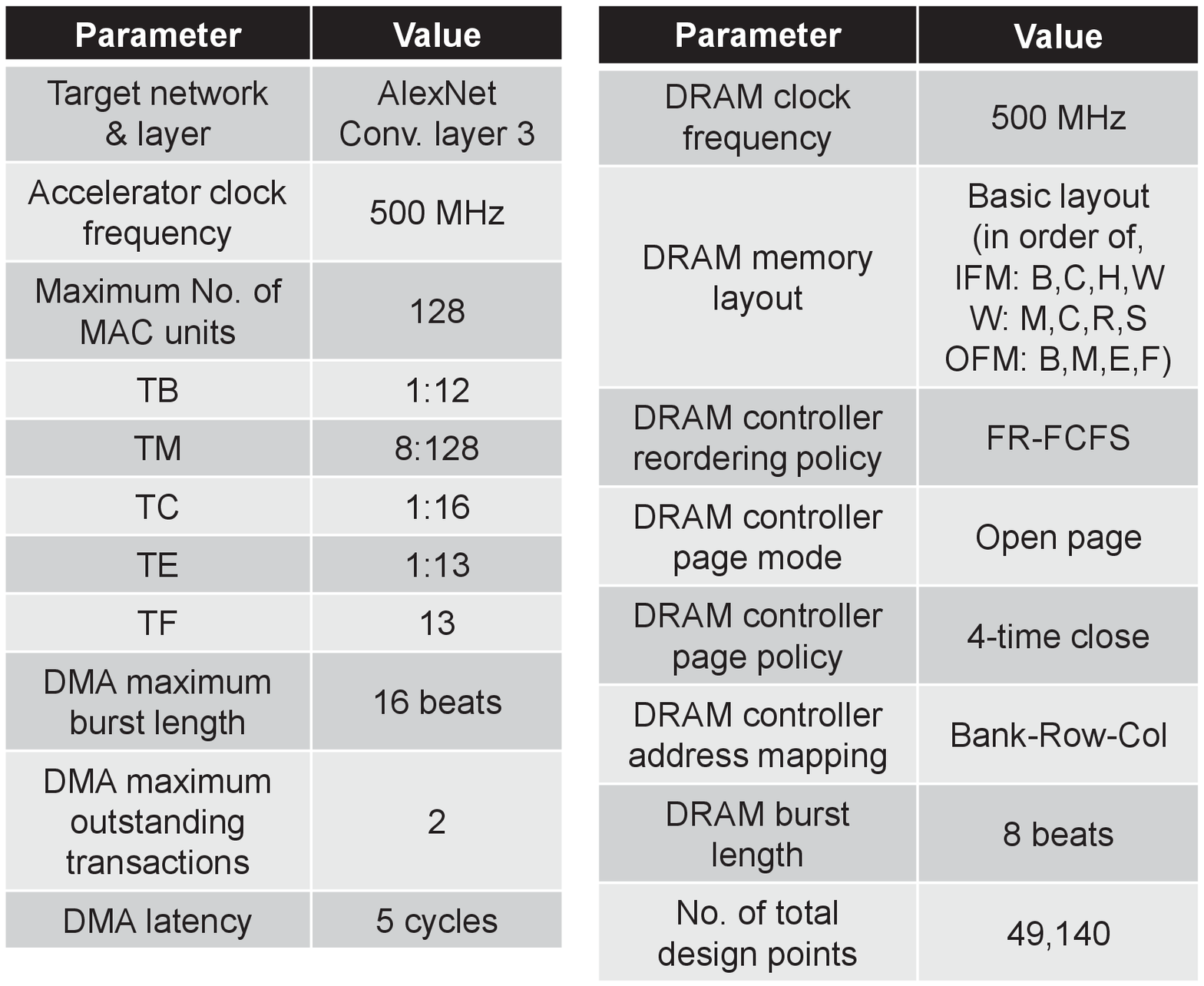}
\end{center}
%\captionof{lstlisting){-}
\caption{Design space for CNN accelerator design}
\label{lst:1}
\end{figure}

\subsection{Performance Estimation Error}

Figure 15 shows that the estimation results of the proposed model match quite well with the simulation results. For the design space in Figure 14, the average estimation error (i.e., the estimation error averaged over the design points) is shown to be as small as 3\%. In contrast, the average estimation error of the conventional model amounts to 36\%. Recalling that a single pixel or weight may take more than one bus cycles to be transferred, it is possible to penalize the DRAM latency by empirically scaling up the communication time [3]. If the scaling factor is identical for all data types (1.75 cycles per pixel), the average estimation error goes down to 9.9\%. If the scaling factor is not identical (1.6, 2.7 and 1.0 cycles per pixel for input feature maps, filters and output feature maps, respectively), the average estimation error goes down to 8\%. Here the scaling factors are empirically chosen to minimize the average estimation error. However, as will be shown in Figure 17 (a), such an empirical scaling simply ends up misleading the accelerator design severely.

\begin{figure}[t] %%% t: top, b: bottom, h: here
\begin{center}
\includegraphics[width=1.0\linewidth]{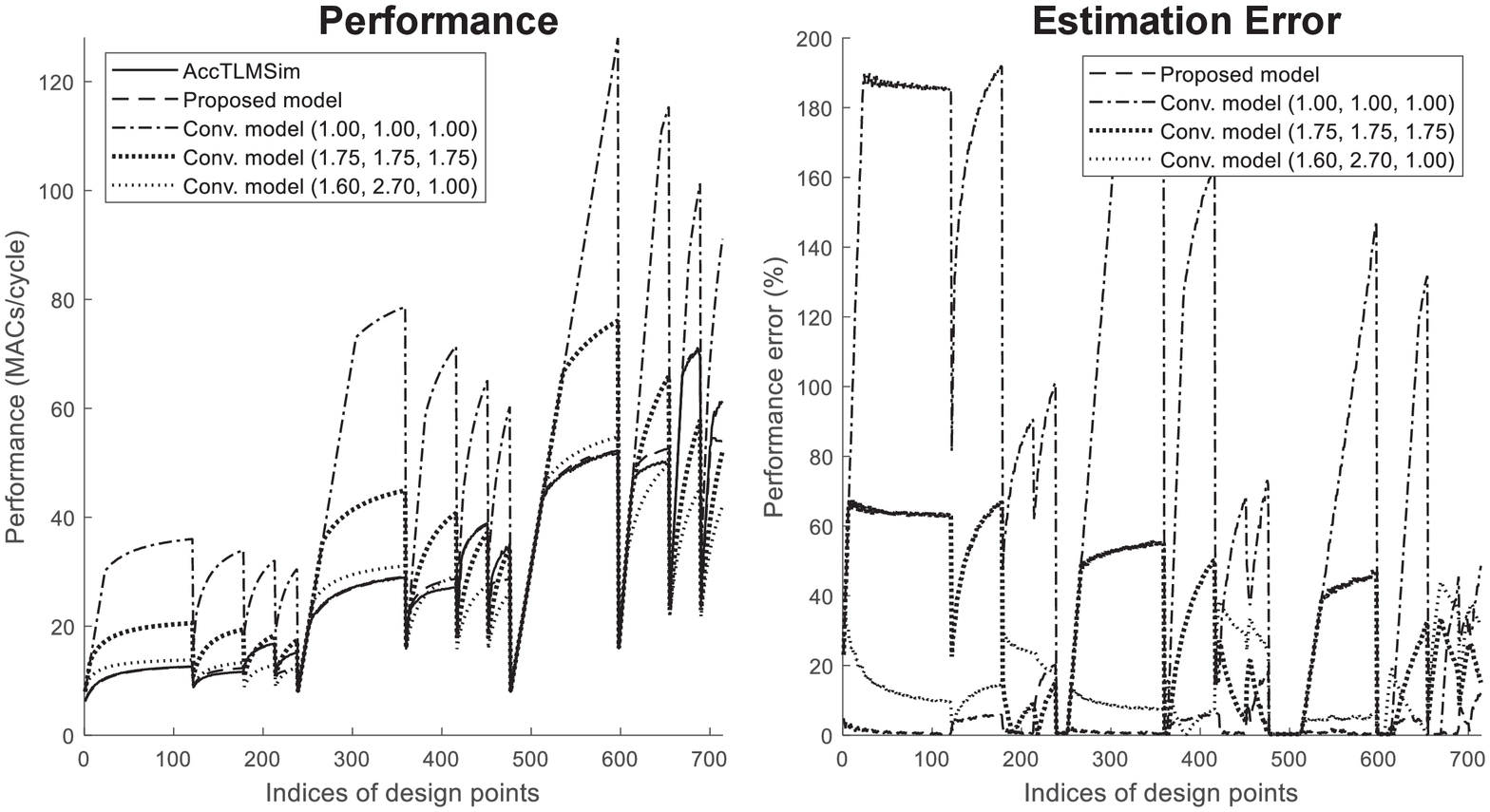}
\end{center}
%\captionof{lstlisting){-}
\caption{Estimation of CNN accelerator performance}
\label{lst:1}
\end{figure}

\begin{figure}[t] %%% t: top, b: bottom, h: here
\begin{center}
\includegraphics[width=1.\linewidth]{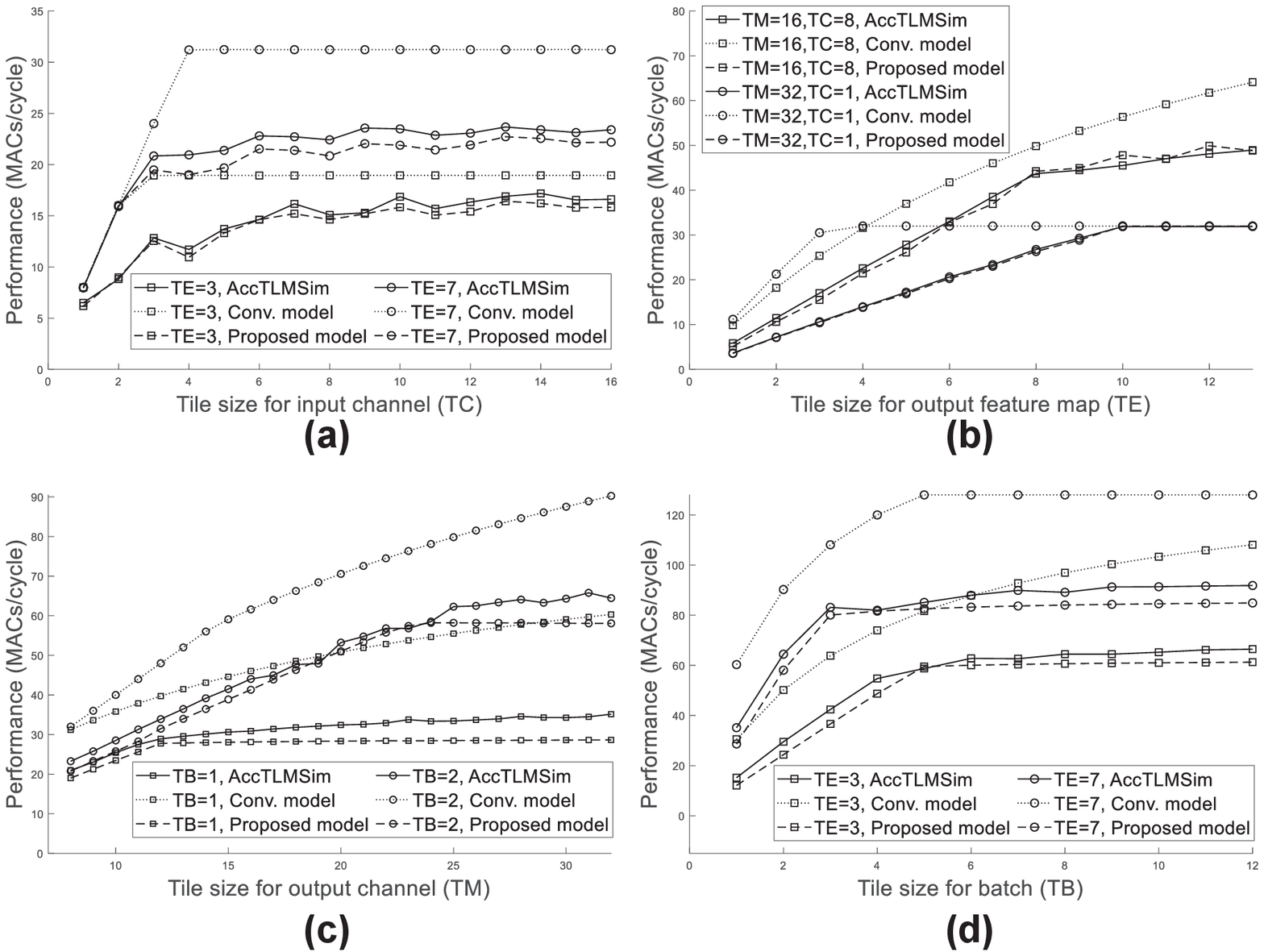}
\end{center}
%\captionof{lstlisting){-}
\caption{Impact of tile sizes on accelerator performance (a) input channel, (b) output feature map, (c) output channel and (d) batch}
\label{lst:1}
\end{figure}

\subsection{Accelerator Design Optimizations}

Figure 16 the impact of tile sizes on the accelerator performance: the tile sizes for the input channel (TC), output feature map (TE), output channel (TM) and batch (TB). In general, the accelerator tends to perform better as the tile sizes increase. However, for sufficiently large tile sizes, the accelerator performance does not continue to improve with tile sizes. For example, as shown in Figure 16 (a), the performance remains to be constant for a sufficiently large input channel tile since the accelerator is communication-limited and thus both the number of MAC operations and the communication time increase linearly with the tile size. In addition, Figure 16 (b) shows that the performance is limited by the number of MAC units (TC$\times$TM) since the output channel tile  is so large that the accelerator is computation-limited.

As shown in the figure, the conventional model fails to predict the communication-limited performance correctly. The conventional model overestimates the communication bandwidth (i.e., one pixel per cycle) since it does not take into account the DRAM latency and bus protocol overhead. In fact, the conventional model is often confused between the computation-limited case and the communication-limited case. For example, as shown in Figure 16 (c), the conventional model assumes that the accelerator is computation-limited for the output channel tile below 15 (i.e., TM \texttt{<} 15) although it is indeed communication-limited (regardless of output channel tile). This confusion causes severe misleading design, as will be shown later.

In contrast, it is obviously found in Figure 16 that the proposed model can predict the actual performance more accurately. As shown in Section IV, the proposed model takes into account the impact of DRAM latency and bus protocol overhead on the communication bandwidth, as opposed to the conventional model that assumes a communication bandwidth of unity (i.e., one pixel per cycle). The estimation error occurs partly due to either the periodic DRAM refresh or the timing uncertainty of the processor core, but it is relatively small, as shown in the figure.

\begin{figure}[t] %%% t: top, b: bottom, h: here
\begin{center}
\includegraphics[width=1.0\linewidth]{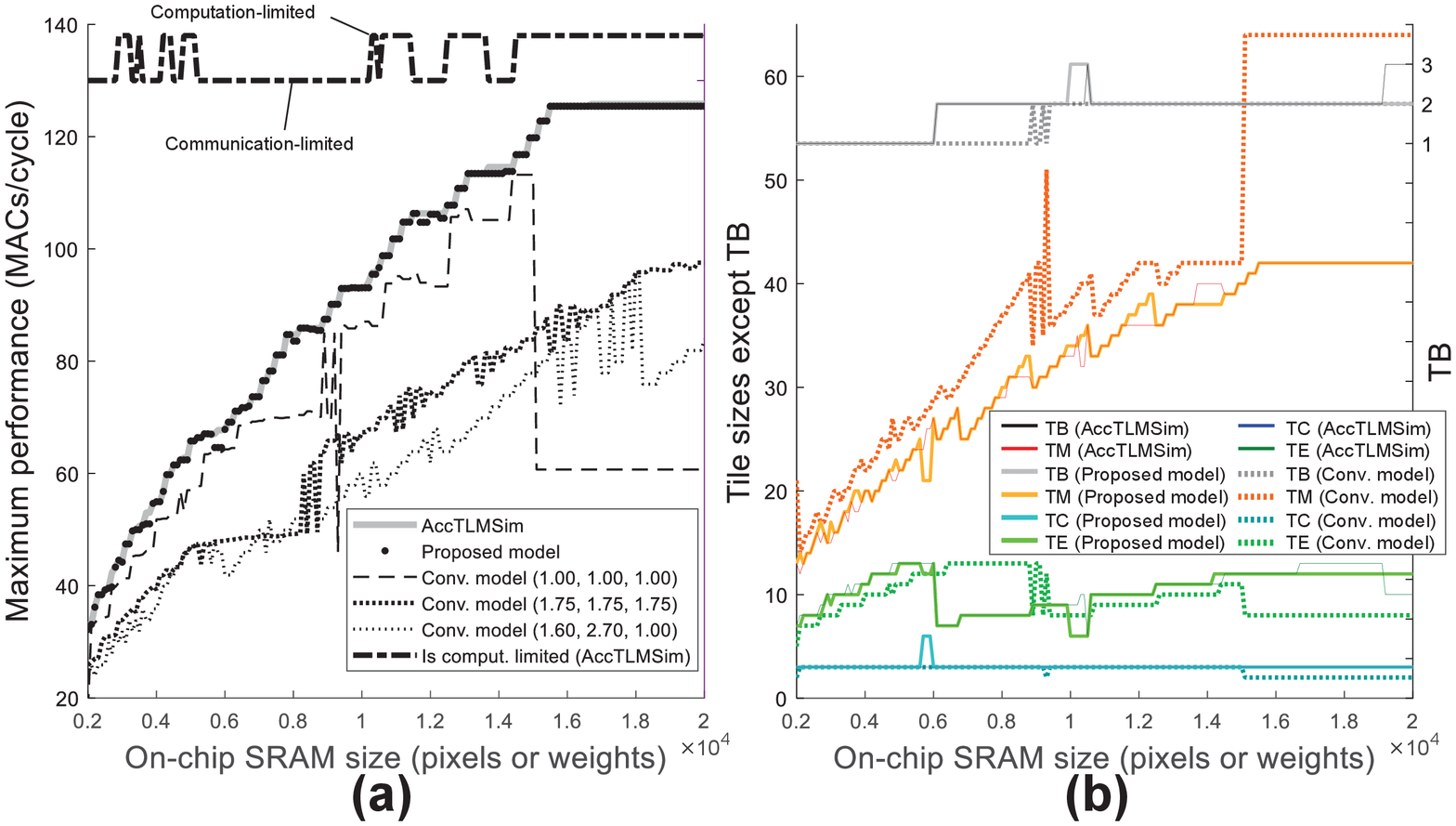}
\end{center}
%\captionof{lstlisting){-}
\caption{Design space exploration with constrained on-chip SRAM size:
(a) maximum performance and (b) the corresponding tile sizes}
\label{lst:1}
\end{figure}

\subsection{Design Space Exploration}

Figure 17 (b) shows the results of the design space exploration whose objective is to maximize the accelerator performance under the constraints of on-chip SRAM size. Noting that the on-chip SRAM size depends on the tile sizes, it follows that the accelerator performance is maximized with respect to tile sizes. If the performance is maximized based on the simulation results, it is clearly shown that the maximum performance always improves with the on-chip SRAM size. The figure also shows that the design point achieving the maximum performance is often communication-limited. Instead of the simulation results, it is possible to maximize the performance based on either of the performance estimation models. Figure 17 shows that the proposed model can serve as a rough approximation of the simulator. On the contrary, the conventional model misleads the accelerator design severely: the performance does not even improve with the on-chip SRAM size. The conventional model may be confused between the computation-limited case and the communication-limited case and thus experience serious performance degradation, especially with the on-chip SRAM size larger than 15k pixels/weights. Figure 18 compares the tile sizes achieving the maximum performance. It is clearly shown that the tile sizes chosen by the proposed model do not deviate significantly from the optimum tile sizes, i.e. those chosen by the accelerator simulator. Moreover, Figure 17 shows that the empirically chosen scaling factors in Figure 15 simply ends up degrading the performance significantly. This implies that the communication bandwidth is not proportional to the communication amount at all.

In order to speed up the design space exploration, either the performance estimation model can be incorporated into the accelerator simulator. Specifically, the design space is first filtered based on the performance estimation model (top-1\% or top-0.1\%), instead of exploring the whole design space, and then, based on the simulation results, the filtered design space is explored to maximize the performance As shown in Figure 18, the proposed model with the top-0.1\% filtering guarantees sufficiently small estimation error and is even more accurate than the conventional model with the top-1\% filtering. In other words, the proposed model can speed up the design space exploration by a few orders of magnitudes by narrowing the design space down.

\begin{figure}[t] %%% t: top, b: bottom, h: here
\begin{center}
\includegraphics[width=1.0\linewidth]{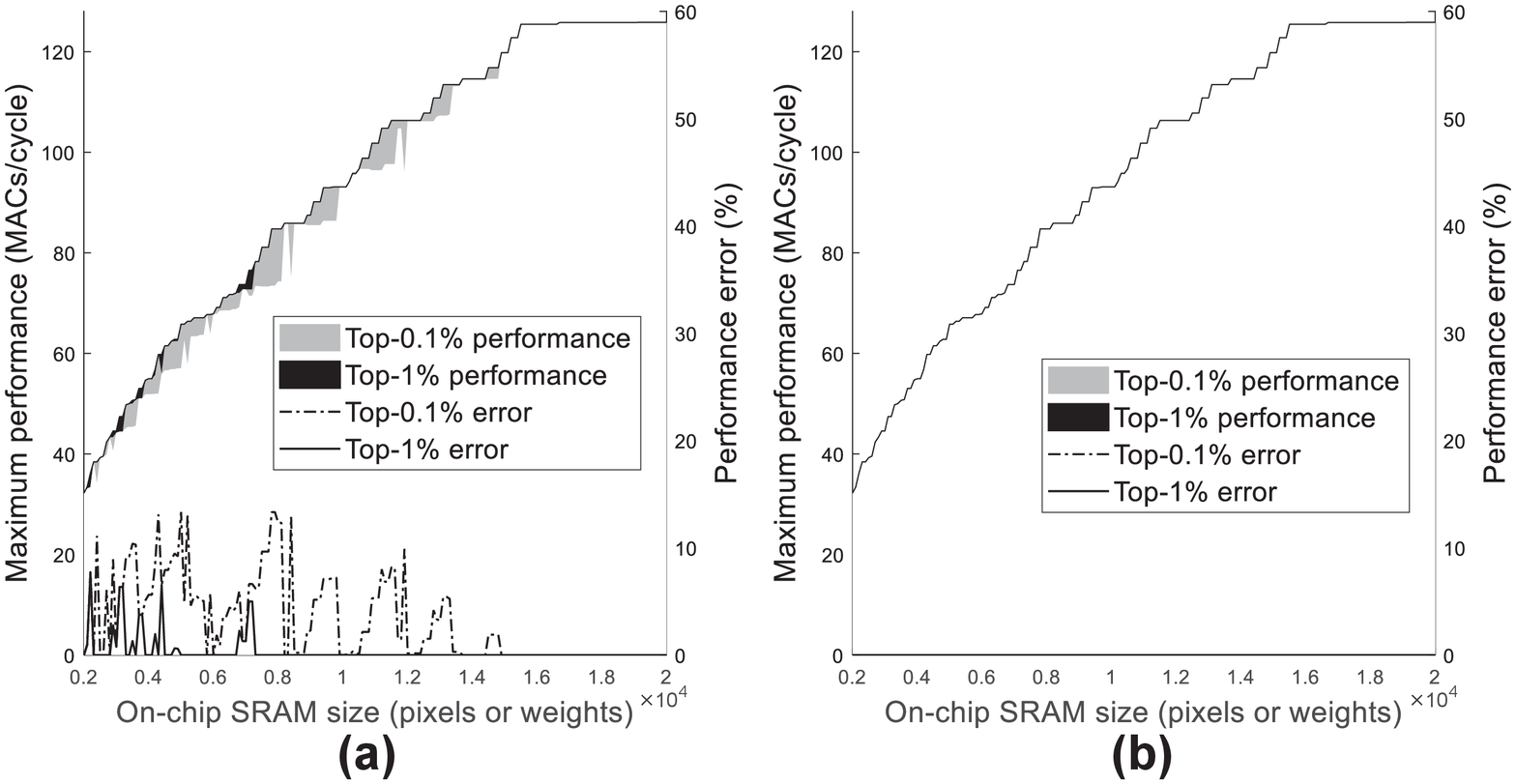}
\end{center}
%\captionof{lstlisting){-}
\caption{Design space exploration combined with performance estimation model: maximum performance and estimation error of
    (a) conventional model and
    (b) proposed model}
\label{lst:1}
\end{figure}

\begin{figure}[t] %%% t: top, b: bottom, h: here
\begin{center}
\includegraphics[width=1.0\linewidth]{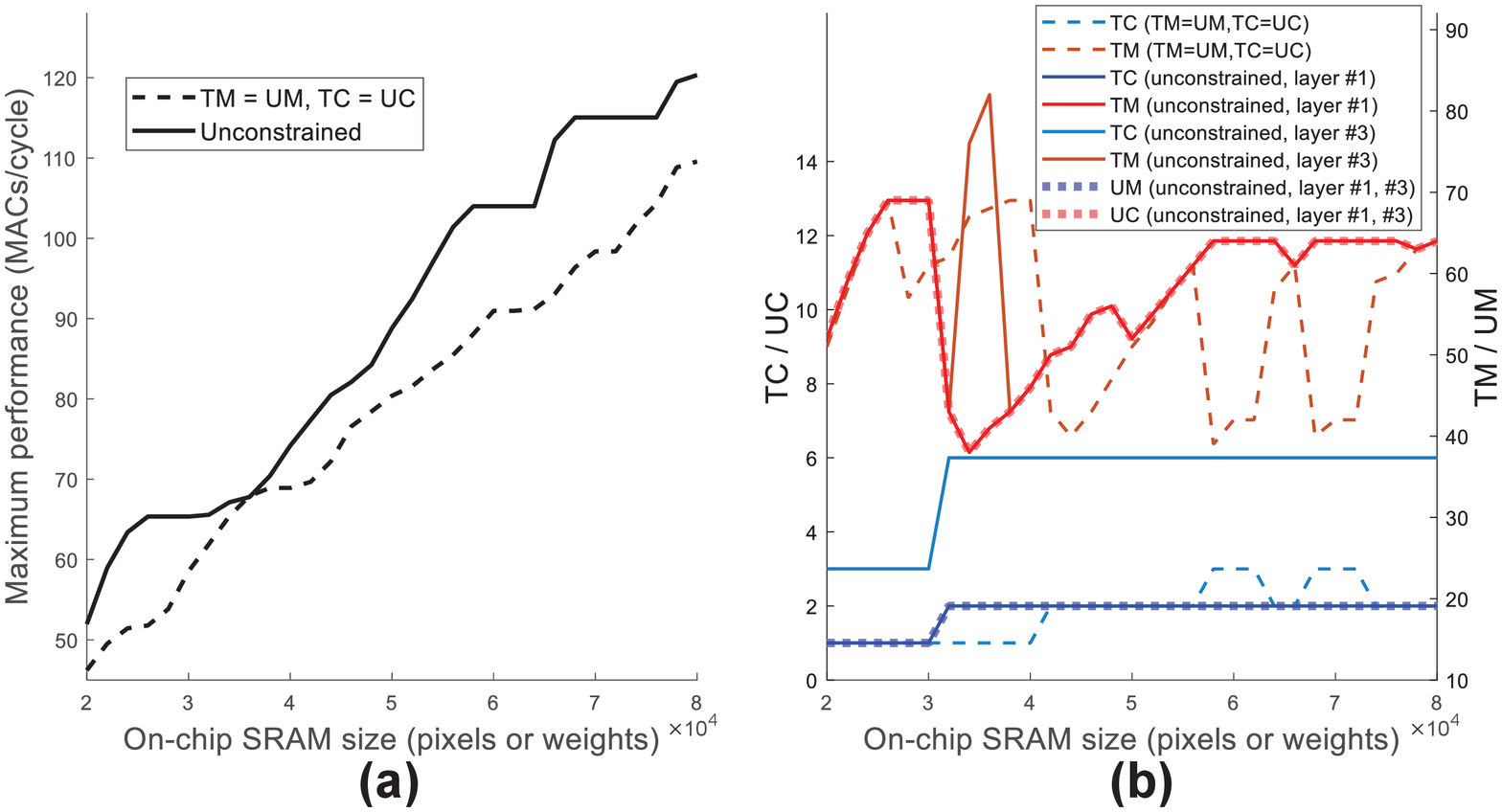}
\end{center}
%\captionof{lstlisting){-}
\caption{Extensions to multiple convolutional layers.
    (a) Maximum performance for the given on-chip SRAM size
    (b) Tile sizes and unroll factors achieving the maximum performance}
\label{lst:1}
\end{figure}

%\begin{figure}	
%	\begin{subfigure}{.5\textwidth}
%		\centering
%		\includegraphics[width=3.5in]{Fig19.eps}
%		\caption{}\label{fig:1a}		
%	\end{subfigure}
%    \newline
%	\begin{subfigure}{.5\textwidth}
%		\centering
%		\includegraphics[width=3.5in]{23\_2.eps}
%		\caption{}\label{fig:1a}		
%	\end{subfigure}
%	\caption{Design space exploration combined with performance estimation model:
%    maximum performance and estimation error of
%    (a) conventional model and
%    (b) proposed model}\label{fig:1}
%\end{figure}

\subsection{Extensions to Multiple Convolutional Layers}

Up to now, it has been assumed in Listing 2 that the unroll factors UM and UC are constrained to be equal to the tile sizes TM and TC, respectively. As mentioned in Section II, this constraint makes it impossible to optimize the tile sizes of a convolutional layer independently of those of the other convolutional layers, thereby limiting the design space. However, the so-called unconstrained loop tiling [11], i.e., the loop tiling whose tile sizes are optimized independently of the unroll factors, provides a potential to utilize the on-chip SRAM size fully.

The extension of the aforementioned design space exploration into multiple convolutional layers may end up with a prohibitively large design space. However, thanks to the proposed model, it is feasible to optimize the tile sizes for multiple convolutional layers. For example, it takes the proposed performance estimation model only a few tens of minutes to explore the design space of millions of design points for the 1st and 3rd convolutional layers of AlexNet. Figure 19 (a) shows that the unconstrained loop tiling improves the maximum performance by up to 25.5\%. This performance gain can be explained by the fact that the unconstrained loop tiling allows the tiling sizes to deviate from the unroll factor (as shown in Figure 19 (b)) and thus exploit the per-layer optimizations of tile sizes.

\section{Conclusion}

In this work, a pre-RTL cycle-accurate accelerator simulator using SystemC TLM was newly proposed for CNN accelerators. The accelerator simulator makes it possible to evaluate the communication bandwidth accurately by taking into the DRAM latency and bus protocol overhead. Using the simulator, the loop tiling was optimized to maximize the performance for the given on-chip SRAM size. In addition, a new performance estimation model was proposed to speed up the design space exploration by a few orders of magnitudes while improving the accuracy significantly. It was applied to the optimizations of the loop tiling of the CNN accelerator, for example, for the unconstrained loop tiling of multiple convolutional layers. Finally, it is worth mentioning that the proposed accelerator simulator together with the performance estimation model can be generally applied to the design of any other accelerators, particularly when the accelerator performance is limited by the communication bandwidth.

\section{Acknowledgements}

This work was supported by Samsung Research Funding \& Incubation Center of Samsung Electronics under Project Number SRFC\-IT1802\-10.

\bibliographystyle{plain}
\bibliography{references}

\vspace{12pt}

\end{document}